\documentclass[debug]{rmaa} 

\usepackage{paralist}

\usepackage{psfrag,color} 

\usepackage[latin1]{inputenc}

\usepackage{graphics}

\def\msun{\mbox{M$_\odot$}} 

\def\Zsun{\mbox{$Z_\odot$}}

\title{H~{\sc ii} REGIONS AND THE PROTOSOLAR HELIUM, CARBON, AND OXYGEN
ABUNDANCES IN THE CONTEXT OF GALACTIC CHEMICAL EVOLUTION}

\author{ L. Carigi\altaffilmark{1} and M. Peimbert\altaffilmark{1} }

\altaffiltext{1}{Instituto de Astronom\'{\i}a, UNAM, M\'exico.}

\shortauthor{Carigi \& Peimbert}

\shorttitle{Galactic chemical evolution}

\fulladdresses{ \item Instituto de Astronom\'{\i}a, Universidad Nacional
Aut\'onoma de M\'exico, Apdo. Postal 70-264, M\'exico 04510 D.F., M\'exico
(carigi, peimbert@astroscu.unam.mx).}

\listofauthors{L. Carigi \& M. Peimbert} \indexauthor{Carigi, L.}
\indexauthor{Peimbert, M.}

\resumen{
Presentamos  modelos de evoluci\'on qu\'{\i}mica del disco Gal\'actico
con diferentes rendimientos dependientes de $Z$.
Encontramos que una tasa moderada
de p\'erdida de masa en estrellas masivas de metalicidad solar produce un excelente
ajuste con los gradientes de C/H y C/O del disco de la Galaxia.
El mejor modelo reproduce: las abundancias de H, He, C, y O derivadas de l\'{\i}neas de recombinaci\'on
en M17, las abundancias protosolares y las relaciones
C/O-O/H, C/Fe-Fe/H y O/Fe-Fe/H derivadas de estrellas de la vecindad solar.
El acuerdo del modelo con las abundancias protosolares implica que el Sol se origin\'o a 
una distancia galactoc\'entrica similar a la actual.
El modelo para $r=3$ kpc implica que una fracci\'on de las estrellas en la direcci\'on
del bulbo se form\'o en el disco interno.
Nuestro modelo reproduce la relaci\'on C/O-O/H derivada de regiones H~{\sc ii}
extragal\'acticas en galaxias espirales.
}

\abstract{ We present chemical evolution models  of the Galactic disk
with different $Z$-dependent yields. We find
that a moderate mass loss rate for massive stars of solar metallicity
produces an excellent fit to the observed C/H and C/O gradients of
the Galactic disk.
The best model also fits: the H, He, C, and O abundances derived from
recombination lines of M17, the protosolar abundances,
and the C/O-O/H, C/Fe-Fe/H, and O/Fe-Fe/H relations derived from solar vicinity stars.
The agreement of the model with the protosolar abundances
implies that the Sun originated at a galactocentric distance similar
to the one it has.
Our model for $r=3$ kpc implies that a fraction of the
stars in the direction of the bulge formed in the inner disc.
We obtain a good agreement between our model and
the C/O versus O/H relationship derived from extragalactic H~{\sc ii}
regions in spiral galaxies.
}

\addkeyword{galaxies: abundances} \addkeyword{galaxy: bulge}
\addkeyword{galaxies: evolution} \addkeyword{H~II regions: M17, Orion
nebula } \addkeyword{ISM: abundances} \addkeyword{Sun: abundances}

\begin{document}

\maketitle


\section{Introduction} \label{sec:intro}

The comparison of detailed Galactic chemical evolution models, GCE models,
with accurate abundance determinations of stars and gaseous
nebulae provides a powerful tool to test the chemical evolution
models and the accuracy of observational abundance determinations of
stars of different ages and of H{\sc~ii} regions located at different
galactocentric distances.

In this paper we will compare our models with stellar and H{\sc~ii}
regions abundances to test if the H{\sc~ii} region abundances derived
from recombination lines agree with the stellar abundances, in particular
with the protosolar abundances that correspond to those present in the
interstellar medium 4.5 Gyr ago.  Also our GCE models can be used to
constrain the C yields for massive stars, the C yield is not well known
and we will vary it to obtain the best fit between our GCE models and
the observational data.

Carigi and Peimbert (2008, hereinafter Paper I) presented chemical
evolution models of the Galactic disk for two sets of stellar yields that
provided good fits to: a) the O/H and C/H gradients (slope and absolute
value) derived from H{\sc~ii} regions based on recombination lines
(Esteban et al. 2005) and including the dust contribution \citep{est98},
and b) the $\Delta Y/\Delta Z$ value derived from the Galactic H{\sc~ii}
region M17, and the primordial helium abundance, $Y_p$ obtained from
metal poor extragalactic H{\sc~ii} regions (Peimbert et al. 2007). In
Paper I, based on our GCE models and combined with the constraints available,
we were not able to discriminate between the stellar evolution models
assuming high wind yields for massive stars, HWY,  and those assuming
low wind yields for massive stars, LWY.

Previous works have focused on the test of stellar yields
using GCE models constrained by chemical gradients obtained by 
different methods, gradients that in general show similar slopes but 
a considerable spread in the absolute O/H ratios, 
e.g. \citet{pra94,car96,chi03a,rom10}. To test the stellar yields 
it is necessary to have good absolute
abundance values of stars and H{\sc~ii} regions.  To determine the O/H
abundances of H{\sc~ii} regions in irregular and spiral galaxies many
methods have been used in the literature. Most of them have been based
on fitting photoionized models to observations or by determining the
electron temperature, $T$, from the 4363/5007 [O III] ratio directly from
observations. A comparison of  many of the different methods used has
been made by \citet{kew08}. They find that the O/H differences
derived by different methods between two
given H{\sc~ii} regions amount to $0.10 - 0.15$ dex.  Alternatively for
all the methods the absolute difference for a given H{\sc~ii} region
is considerably larger reaching values of 0.7 dex for extreme cases
(see Figure 2 in their paper and the associated discussion).  Most of
the differences among the various calibrations are due to the temperature
distribution inside the nebulae. In this paper we will use only abundances
of H{\sc~ii} regions based on recombination lines of H, He, C, and O,
these lines depend weakly on the electron temperature, they are roughly
proportional to $1/T$, therefore the relative abundances among these
four elements are practically independent of the electron temperature.

There are two frequently used methods to derive C and O gaseous abundances
from H{\sc~ii} regions: a) the most popular one based on collisionally
excited lines (or forbidden lines) and the  $T$(4363/5007) temperatures,
the FL method, and b) the one based on C and O recombination lines,
the RL method. The RL method produces gaseous O and C abundances higher
by about 0.15 to 0.35 dex than the FL method.  The RL method is almost
independent of the electron temperature, while the FL method is strongly
dependent on the electron temperature. It is possible to increase the
FL abundances under the assumption of temperature inhomogeneities to
reach agreement with the RL values. The temperature distribution can be
characterized by the average temperature, $T_0$, and the mean square
temperature variation, $t^2$, \citep[e.g.][]{pei67,pei02}. The $t^2$
values needed to reach agreement between the RL and the FL abundances
are in the 0.02 to 0.05 range, while the photoionization models predict
typically $t^2$ values in the 0.003 to 0.01 range, this discrepancy
needs to be sorted out \citep[e.g.][and references therein]{pei11}.

Paper I is controversial because the C/H and O/H gaseous abundances of
the H{\sc~ii} regions have been derived from recombination lines (that is
equivalent to the use of  $t^2 \neq 0.000$ and forbidden C and O lines)
and the assumption that 20\% of the O atoms and 25 \% of the C atoms
are trapped in dust grains (0.08 dex and 0.10 dex respectively). These
assumptions increase the O/H ratio by about 0.25 to 0.45 dex relative
to the gaseous abundances derived from $T$(4363/5007),  the forbidden
O and C lines, and the assumption that $t^2 = 0.00$.

Due to the controversial nature of the  H{\sc~ii} region abundances
used in Paper I and that we were not able to discriminate between the
two sets of stellar yields adopted we decided to test our GCE models
further by including additional observational constrains: a) the Asplund
et al. (2009) protosolar abundances that provide us with the O/H, C/H,
Fe/H, and $\Delta Y/\Delta O$ in the interstellar medium 4.5 Gyr ago,	
b) the C/H, O/H, and Fe/H by \citet{ben06} for young F and G stars of
the solar vicinity, c) the O/H, C/H, and He/H derived from B stars by
\citet{prz08}, and d) throughout this paper for all the H{\sc~ii} regions
we will use abundances derived from recombination lines  and to obtain the
total abundances we will increase the gaseous abundances by 0.10 dex in
C and 0.12 dex in O to take into account the fraction of atoms trapped in
dust grains, with the exception of the metal poor irregular galaxies for
which we will use an 0.10 dex depletion for O \citep{est98,mes09,pea10}.

We also decided to compare our best models with the C/O versus O/H results
derived by  \citet{est02,est09} from bright H{\sc~ii} regions in nearby
spiral galaxies based on recombination lines and to make a preliminary
discussion of a comparison between our models for $r$ = 3kpc and the
stars in the direction of the galactic bulge obtained by \citet{ben10a}
and \citet{zoc08}.

The symbols  $C$, $O$, $X$, $Y$, and $Z$ represent carbon, oxygen,
hydrogen, helium, and heavy element abundances by unit mass respectively;
while C/H, O/H, Fe/H,  C/O, C/Fe, and O/Fe represent the abundance ratios
by number.

In Section~\ref{sec:models} we discuss the general properties
of the chemical evolution models, we discuss infall models for
the Galaxy with two sets of stellar yields, the HWY and the LWY. In
Section~\ref{sec:gradient} we show the prediction of the current abundance
gradients for the interstellar medium (ISM) of the Galactic disk and
compare them with Galactic H{\sc~ii} region abundances derived from
recombination lines that include the dust correction, in addition for the
solar vicinity we present the chemical history of the ISM and compare
it with the chemical abundances of stars of different ages; we define
the solar vicinity as a cylinder perpendicular to the galactic plane,
centered in the Sun, with a radius of 0.5 kpc, that extends into the halo
to include the stars in the cylinder.  In Section~\ref{sec:M17} we compare
our chemical evolution models with the protosolar chemical abundances
and with those of the H{\sc~ii} region M17. Based on the comparison
between the observations and the models in Section~\ref{sec:intermediate}
we present a new Galactic chemical evolution model, with intermediate
mass loss due to interstellar winds, IWY, that produces considerably
better adjustments with the observations. In Section~\ref{sec:other}
we compare the IWY model with additional observations, those provided
by extragalactic H{\sc~ii} regions in spiral galaxies, and those
provided by stars in the direction of the Galactic bulge.  The conclusions are presented in
Section~\ref{sec:conclusions}.  A preliminary account of some of the
results included in this paper was presented elsewhere \citet{pei10}.


\section{Chemical Evolution Models With High Wind Yields and Low Wind
Yields} \label{sec:models}

We present chemical evolution models for the Galactic disk.  The models
have been built to reproduce the present gas mass distribution, (see
Fig. 1, left panel) and the present-day O/H values for H~{\sc ii}
regions in the Galaxy  for $6<r$(kpc)$<11$, (see Fig. 1, upper right
panel) listed by \citet{gar07}. In what follows we describe in detail
the characteristics of the models.

i) In the models the halo and the disk are projected onto a single disk
component of negligible width and with azimuthal symmetry, therefore
all functions depend only on the galactocentric distance $r$ and time $t$.

ii) The models focus on $r \ge 4$ kpc, because the physical processes
associated with the Galactic bar are not considered.

iii) The models are based on the standard chemical evolution equations
originally written by \citet{tin80} and widely used to date. See e.g.
\citet{pag09}, \citet{mat00}, and \citet{pra08}.

iv) The age of the models is 13 Gyr, the time elapsed since the beginning
of the formation of the Galaxy.

v) The models are built based on an inside-out scenario with infalls
that assume primordial abundances ($Y_p=0.2477$, $Z=0.00$, Peimbert et
al. 2007). The adopted double infall rate is similar to that presented by
\citet{chi97}, as a function of $r$ and  $t$, and  is given by $IR(r,t)
= A(r) e^{-t/\tau_{\rm halo}} + B(r) e^{-(t-1 Gyr)/\tau_{\rm disk}}$,
where the halo formed in the first Gyr with a timescale, $\tau_{\rm halo}
= 0.5 $ Gyr, and the disk started forming immediately after  with longer
timescales that depend on $r$, $\tau_{\rm disk}=(8 \times r/r_\odot -
2)$ Gyr. We adopt 8 kpc for the galactocentric distance of the solar
vicinity, $r_\odot$.   The variables $A(r)=10 \msun pc^{-2}\times
e^{-(r(kpc)-r_\odot)/3.5kpc}$ and $B(r)=40  \msun pc^{-2}\times
e^{-(r(kpc)-r_\odot)/3.5kpc}$ are chosen to match the present-day mass
density of the halo and disk components in the solar vicinity that
amount to 0.5 and 49.5 \msun $pc^{-2}$, respectively, where the mass
lost by the stars and the halo gas have been incorporated into the disk.
Moreover with the $A(r)$ and $B(r)$ variables the models reproduce the
radial profile of the total mass in the Galaxy, $M_{tot}(r)= 50 \msun
pc^{-2}\times e^{-(r(kpc)-r_\odot)/3.5kpc}$ \citep{fen03}.

vi) The models assume the Initial Mass Function (IMF) proposed by
\citet{ktg93}, in the mass interval given by $0.08 - 80.0$ \msun \ mass
range for $Z > 10^{-5}$, and in the 9.0-80.0 \msun \ mass range for $Z <
10^{-5}$. We consider that Population III do not include objects with
less than 9.0 \msun \ and that the change from Pop III to Pop II.5 occurs
at $10^{-3.5}Z_\odot$ \citep{ake04,bro04}.

vii) The models include a star formation rate that depends on time
and galactocentric distance, $ SFR(r,t) = \nu  M^{1.4}_{gas}(r,t)
\ (M_{gas}+M_{stars})^{0.4}(r,t)$, taken from \citet{mat89} and
\citet{mat99}, where $\nu$ is a constant in time and space.  This SFR
formula considers the feedback between gas and stars.  In our models $\nu$
is chosen in order to reproduce the present-day radial distribution of
gas surface mass density.  We adopted $\nu$ values of 0.019 and 0.013
for the HWY and the LWY models respectively. We assumed a $\nu$ value 5
times higher during the halo formation than that adopted for the disk.
These $\nu$ values combined with the infall rate adopted reproduce the
chemical abundances shown by halo and disk stars (see Fig. 2).

viii) The only difference between the HWY and the LWY sets is the assumed
mass-loss rate due to stellar winds by massive stars with  $Z = 0.02$,
see Figure 1  of \citet{car08}.  All stellar
yields are metal dependent. We interpolate and extrapolate linearly
the yields by mass and metallicity.

The HWY set includes:

A) For massive stars (MS), those with $8 < m/\msun < 80$, the yields
by: a) \citet{hir07} for  $Z = 10^{-8}$ (with rotation velocity between
500 and 800 km/s, depending on stellar mass); b) \citet{mey02} for  $Z
= 10^{-5}$ and $Z = 0.004$ (with rotation velocity = 300 km/s); c)
\citet{mae92} for $Z = 0.02$ (high mass-loss rate with no rotation
velocity); d)  Since Fe is a chemical element used as observational
constraint in most of the chemical evolution models and Fe yield is not
computed by \citet{hir07}, \citet{mey02}, and \citet{mae92}, we adopt
\citet{woo95} only for the Fe yields (Models B, for 12 to 30 \msun;
Models C, for 35 to 40 \msun; while for $m > 40  \msun$ , we extrapolated
the $m = 40 \msun$ Fe yields, following Carigi \& Hernandez 2008 ).
Our models, that include the Fe yields and the O yields by \citet{mey02},
can reproduce the O/Fe-Fe/H trend of the halo stars (see  Fig 2).

B) For low and intermediate mass stars (LIMS), those with $0.8 \leq
m/\msun \leq 8$, we have used the yields by \citet{mar96,mar98} and
\citet{por98} for the $Z=0.004$ to $Z=0.02$ range.

C)  For Type Ia SNe we have used the yields by \citet{thi93} in the
the SNIa formulation of \citet{gre83}.  A fraction, $Abin$, of the binary
stars of LIMS  with a total mass between 3 and 16 \msun \ are progenitors
of SNIa.  We  assumed $Abin= 0.066$    for HWY model, while $Abin= 0.094$
for LWY models.  These  fractions are needed to reproduce the age-[Fe/H]
relation of the disk stars, the O/Fe-Fe/H relation of the disk stars
of solar vicinity, and the protosolar Fe/H value for each model (see
Fig. 2).  

In the LWY set we have updated the yields of massive stars only for  $Z
= 0.02$ assuming the yields by \citet{hir05} with rotation velocity =
300 km/s.  The rest of the stellar yields are those included in the high
wind set.

ix)  For MS the models include the stellar lifetimes given by
\citet{hir07}, \citet{mey02}, and \citet{hir05}, while for LIMS include
the main sequence lifetimes from \citet{sch92}.

x)  The models do not consider any type of outflows from the Galaxy due
to its deep potential well. In addition, we discard outflows of C and O
rich material from SNe, due to the small C/O ratios observed in H~{\sc
ii} regions. See \citet{car95,car99}.

xi) We do not include radial flows of gas or stars.

Since the solar vicinity and the Galactic disk contain stars and H~{\sc
ii} regions of a broad range of metallicities, our galaxy is a proper
laboratory to study the $\Delta Y/\Delta O$ and $\Delta C/\Delta O$
behavior at high $Z$ values and to  test observationally the predictions
of the HWY and the LWY models. Furthermore the main elements affected
by the winds of massive stars are He, C, and O, therefore a careful
comparison of the abundances predicted by the chemical evolution models
with the observed values will permit to constrain the mass loss rate
and consequently the yields of massive stars.


\section{The Galactic H~{\sc ii} Regions Gradients and the Chemical
History of the Solar Vicinity } \label{sec:gradient}

\begin{figure}[!t] \includegraphics[width=\columnwidth]{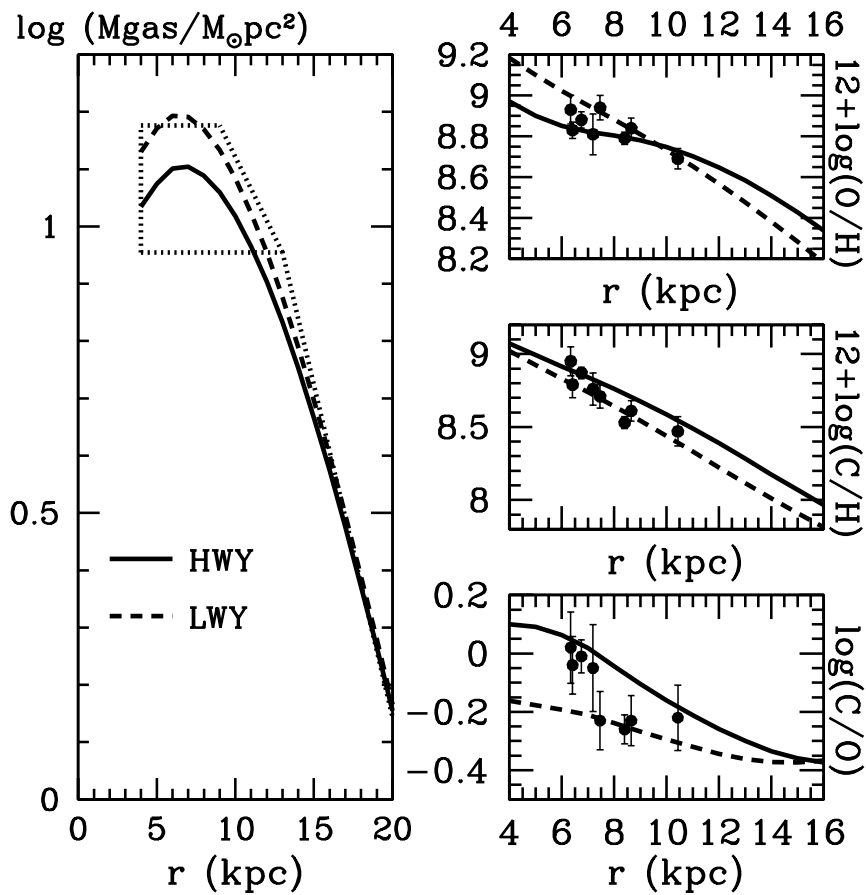}
\caption{ Present-day radial distribution of:
 gas surface mass density ({\it left panel}), and ISM abundance ratios
({\it right panels}).  Predictions of our chemical evolution models for
the Galactic disk at the present time: HWY ({\it continuous lines}),
LWY ({\it dashed lines}).  Observational data: {\it area enclosed by
dotted lines}: average gas surface density distribution by \citet{kal09},
{\it filled circles}: H~{\sc ii} regions, gas \citep{gar07} plus dust
values (see text).} \label{fig:yields} \end{figure}

In Figure 1  we show the present-day radial distribution of the gaseous
mass and the O/H, C/H, and C/O gradients in the Galactic disk predicted
by the HWY and LWY models.

For Figure 1 we have chosen observational constraints that represent
the current gas mass and the abundances of the Galactic disk.  We have
taken the average surface density as a function of the galactic radius
derived from H~{\sc i} shown by \citet{kal09} in their Figure 5.
We have assumed the abundance ratios of the gas component determined
from H~{\sc ii} regions based on recombination lines by \citet{gar07}
corrected by dust depletion. It should be noted that the the H~{\sc ii}
regions used by \citet{gar07} are high density young objects that do
not contain WR stars, consequently they have not been polluted by the
evolution of their ionizing stars and their O/H and C/H values are
representative of the present value of the galactic interstellar medium.

Based on Figure 1, it can be noted that both models successfully
reproduce the current radial distribution of gas surface mass density
and the C/H and O/H gradients at the one $\sigma$ error but neither
of them reproduces the C/O gradient for all Galactocentric distances.
The HWY model adjusts the C/O values of H~{\sc ii} regions for  $r<7.3$
kpc, while the LWY model does so for $7.5 <r$(kpc)$< 9$.

In the literature there are chemical abundance determinations
from recombination lines of Galactic H~{\sc ii} regions only for
$6<r$(kpc)$<$11.  At present, the evidence for the flattening of the
Galactic gradients is not conclusive: \citet{vil96}  based on H~{\sc ii}
regions found a flat O/H gradient for $r > 14$ kpc, while more recent
work based on Cepheids did not confirm this result, see \citet{mats10}.
Additional observations of high quality obtained with the same method
are needed to established the behavior of the gradients for $r >
11$ kpc.  With an inside-out scenario without dynamical effects it
is difficult to reproduce a possible flattening of the O/H gradient,
see \citet{mats10}.   \citet{ces07} in an inside-out scenario
reproduce the flattening of radial gradients shown by Cepheids for
$r > 7$ kpc assuming a constant surface density for the halo.  In our
models it is equivalent to adopt a constant $A(r)$, which means that the
volumetric density of the halo increases towards the outer parts of the
Galaxy, a density distribution that we consider unlikely.  According to
\citet{ros08} and \citet{san09}, the flattening is a consequence of
stellar migration and of a break in the star formation rate at large
radii. A third possibility is that the formation of the Galaxy had a
modest stellar formation previous to the inside-out scenario, support
for this idea comes from the increase of the average age of the stars at
large galactocentric distances \citep[][and references therein]{vla11}.  

\begin{figure}[!t] \includegraphics[width=\columnwidth]{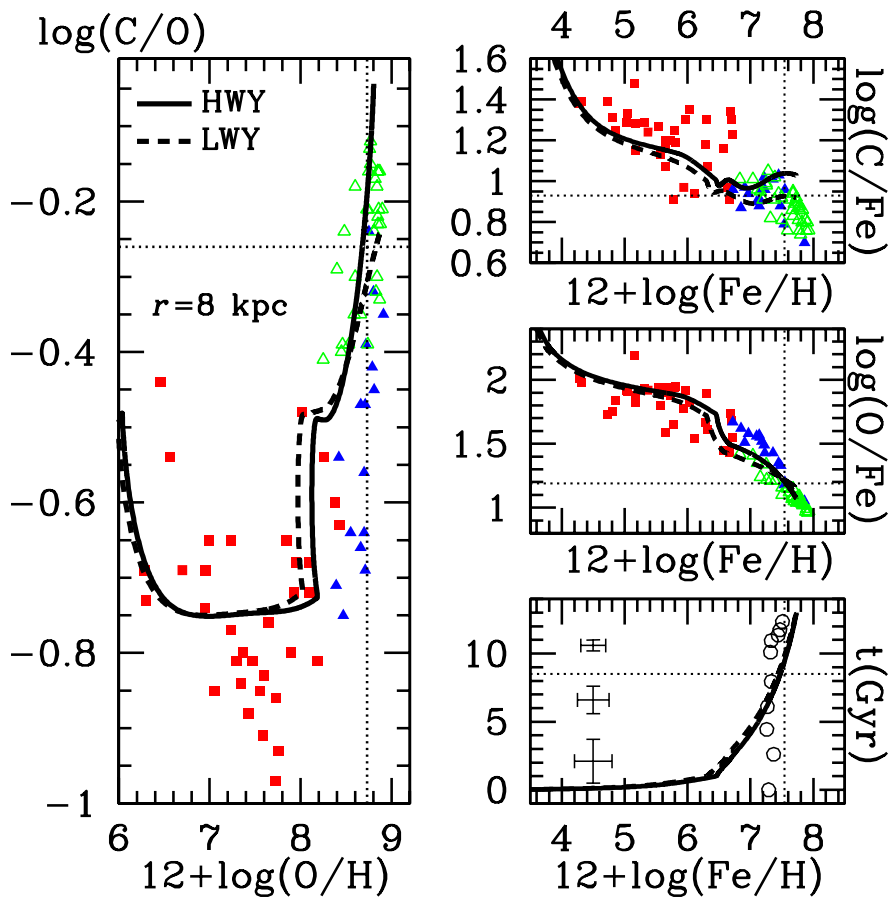}
\caption{
 Chemical evolution models for the solar vicinity ($r=8$ kpc): HWY ({\it
continuous lines}), LWY ({\it dashed lines}).  The left panel shows the
C/O evolution with O/H.  The right panels show the evolution of C/Fe and
O/Fe with Fe/H, and the Fe/H-time relation.  {\it Filled red squares}:
halo dwarf stars from \citet{ake04}.  {\it Filled blue and empty green
triangles}: thick-disk and thin-disk dwarf-stars from \citet{ben06}.
{\it Dotted lines}: protosolar values from \citet{asp09}.  {\it Open
circles}: mean ages and Fe/H values of disk stars with better determined
ages from \citet{nor04}.  {\it Horizontal bars}: internal dispersions
in Fe/H shown by the sample.  {\it Vertical bars}: age average errors.}
\label{fig:solarvinicityHLWY} \end{figure}

In Figure 2 we show the evolution of C/O-O/H,  C/Fe-Fe/H, O/Fe-Fe/H,
and time-Fe/H, relations, predicted by the HWY and LWY models for $r =
8$ kpc.  The time vs Fe/H plot is an equivalent representation to the
known age-$Z$ relation of the solar vicinity.

For Figure 2 we have chosen observational constraints that represent
the chemical history of the solar vicinity.  We have taken dwarf stars
of the Galactic halo by \citet{ake04}, as representative of the first
Gyr of the evolution, and dwarf stars of the Galactic thick and thin
disk by \citet{ben06}, as representative of the last 12 Gyrs of the
evolution. Also in Figure 2 we present the mean ages and Fe/H values
of disk stars with better determined ages presented by \citet{nor04}
in the lower part of their Figure 28. Moreover, we have assumed the
protosolar abundances  by \citet{asp09}, as representative of $t=8.5$ Gyr.

We converted the chemical abundances determined by \citet{ben06}  to
abundance ratios by number assuming their own solar abundances, see
their paper for references.  Also, we normalized the stellar ages by
\citet{nor04} to the age of the model (13.0 Gyr).

Based on Figure 2, it can be noted that both models produce a reasonable 
fit to the C/O-O/H, C/Fe-Fe/H, and O/Fe-Fe/H trends in the solar vicinity.
 From the C/O-O/H and C/Fe-Fe/H relations it can be seen that the
HWY model predicts more C than observed in metal rich disk stars while
the LWY model predicts less than the HWY.  Alternatively both models
predict a C/Fe plateau for Fe/H higher than solar, while metal rich stars
of the thin disk show a C/Fe decrease.  This observed trend could be
explained if massive stars of $Z>\Zsun$ are less efficient producing C
than massive stars of solar metallicity.  Stellar yields of MS and LIMS
for $Z>\Zsun$ are needed to have a complete picture of the evolution at
high $Z$ \citep{car08b}.

\citet{ces09} and \citet{rom10} focused on the C/O-O/H relation shown by thin disk stars
of the solar vicinity and found that the C/O rise at high O/H values
can be explained partially with metallicity-depend stellar winds in
massive stars \citep{mae92,mey02}.  This result is in agreement with
our previous conclusions \citep[][and Paper I]{car94,car96,car00,car05},
and with those by \citet{pra94}.

The fit of our models to the time-Fe/H relation shown by disk stars of
the solar vicinity is reasonable for $4 <t$(Gyr)$ < 13$, but our models
cannot reproduce the mean behavior of older stars of the Galactic disk.
Probably the reason is that some of these stars originated closer to the
center of the Galaxy and migrated outwards, or belonged to satellite
galaxies with different chemical histories that were captured by our
galaxy.  Most of the stars of the sample have ages lower than 9 Gyr,
corresponding to $t > 4$ Gyr, and  the internal dispersion in Fe/H shown
by old disk stars is higher than that by young stars.  Both models in
the $t < 1$ Gyr range adjust the Fe/H values of the halo stars with ages
between 12 and 13 Gyr.

The main difference between the HWY and the LWY sets is due to the stellar
yields assumed for massive stars at $Z=0.02$. The HWY assume a relatively
high mass-loss rate for massive stars with $Z=0.02$ \citep[yields by]
[]{mae92}, while the LWY assume a relatively low mass-loss rate for
massive stars with $Z=0.02$ \citep[yields by][]{hir05}.
Since the mass loss rate is proportional to the stellar metallicity, the
efficiency of this rate increases with metallicity and becomes important
at $Z \sim \Zsun$.  According to \citet{hir05} mass loss rates are a key
ingredient for the yields of massive stars and the rates assumed by them
are 2-3 smaller than those by \citet{mae92}. This difference between a
high and a low mass-loss rate produces opposite differences in the $C$ and
$O$ yields, these differences occur in the pre-SN and in the SN phases.
In the pre-SN phase massive stars are able to process He into C, and
their stellar winds take away a lot of new C.  Since the O production
happens deeper than  the C production, the wind contribution to the O
yield is much smaller than to the C yield; consequently C constitutes the
largest fraction of heavy elements ejected during the wind phases. While,
in the SN phase the contribution to the C and O total yields depends
mainly on the mass of the CO core, a small fraction of C in the CO core
remains unmodified and is ejected during the explosion. Alternatively
the O production by the SNe is proportional to an important fraction of
the mass of this core; consequently O constitutes the largest fraction
of heavy elements ejected during the SN stage.  Therefore, when the
initial stellar metallicity is higher, the mass loss rate is higher,
the C yield is higher, the CO core is less massive, and consequently
the O yield is smaller.


\section{The Protosolar and the Present Solar Vicinity Abundances}
\label{sec:M17}

To compare GCE models of $Y$, $C$, and $O$ with observations we
need to use the best abundance determinations available for these
elements. We consider that the two most accurate Galactic $Y$, $C$,
and $O$ determinations are the protosolar values (Asplund et al. 2009)
and the M17 H~{\sc{ii}} region values (Paper I).


\subsection{Protosolar abundances }

\begin{figure}[!t] \includegraphics[width=\columnwidth]{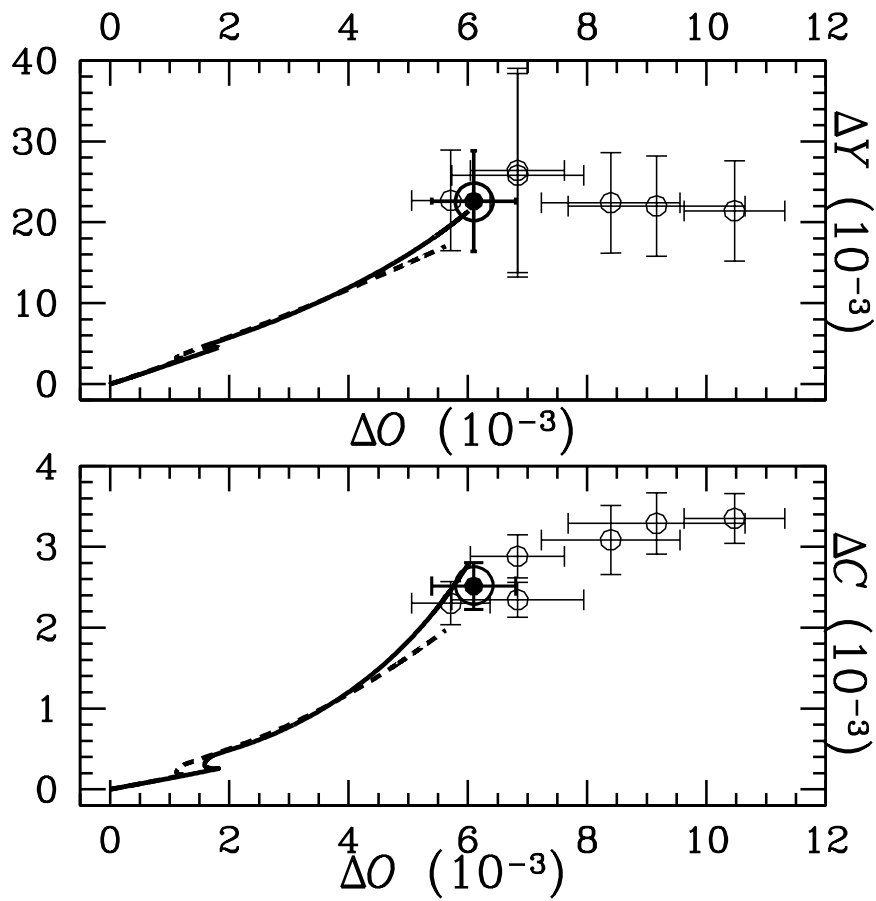}
\caption{
 $\Delta Y$ vs $\Delta O$ ({\it upper panel}) and $\Delta C$ vs
$\Delta O$ ({\it lower panel}) for the solar vicinity ($r$ =  8
kpc). Evolution from $t=0$ ($\Delta Y = \Delta C = \Delta O = 0 $)
to 8.5 Gyr (Sun-formation time) predicted by models that assume HWY
({\it continuous lines}), and LWY ({\it dashed lines}).  ${\odot}$ :
Protosolar values by \citet{asp09}. {\it Open circles:} Protosolar values
from photospheric data by different authors compiled by \citet[][Table
4]{asp09} and corrected for gravitational settling \citep{asp09}.}
\label{fig:HeliumCSunHLWY} \end{figure}

In order to study the time-agreement with the protosolar values, we show
in Figure 3 the predicted evolution of $\Delta Y$ vs  $\Delta O$ and
$\Delta C$ vs $\Delta O$ from 0 to 8.5 Gyr (the Sun-formation time). The
$\Delta Y$  value for Figure 3 (and throughout this paper) is given by
$\Delta Y = Y -Y_p$, where $Y_p$ is the primordial He abundance,
and amounts to 0.2477 \citet{pei07}.  In this figure $\Delta O = O$, and
$\Delta C = C$ because at the time of the primordial nucleosynthesis
O and C are not produced. For $t$ = 0 the models start at $\Delta Y$,
$\Delta O$, and $\Delta C$ equal to zero, and in this figure the evolution
of the models stops at $t$ = 8.5 Gyr, the time the Sun was formed.
Also in this figure we present some of the most popular $Y$, $C$, and
$O$ protosolar abundances of the last 21 years. These photospheric solar
abundances were obtained  by Anders \& Grevesse (1989), Grevesse \& Noels
(1993), Grevesse \& Sauval (1998), Lodders (2003), Asplund, Grevesse \&
Sauval (2005), and Lodders, Palme \& Gail (2009), abundances that were
compiled by \citet{asp09} in their Table 4. To obtain the protosolar
abundances of the heavy elements, the photospheric abundances were
increased by 0.04 dex, amount that takes into account the effect of
gravitational settling. We used the protosolar $X$ values shown in Table 4
by \citet{asp09} to change the abundances by number to abundances by mass.

From Figure 3 it can be noted that: a) the HWY model fits very well
the $Y$ and $O$ protosolar values, while the predicted $C$ is about 1
$\sigma$ error higher than observed, b) the LWY model matches the $Y$
and $O$ protosolar values within 1 $\sigma$, but the predicted $C$ is
about 2 $\sigma$ lower than observed, c) the solar abundances predicted
by our models are in much better agreement with the recent He, C, and
O protosolar values determined by \citet{asp09} than with the previous
ones compiled by them, and d) the He protosolar abundance determinations
have remained almost constant over the years, while the C and O ones
have decreased.


\subsection{$\Delta Y$ vs $\Delta O$ evolution compared with M17 and
young B stars of the solar vicinity}

\begin{figure}[!t] \includegraphics[width=\columnwidth]{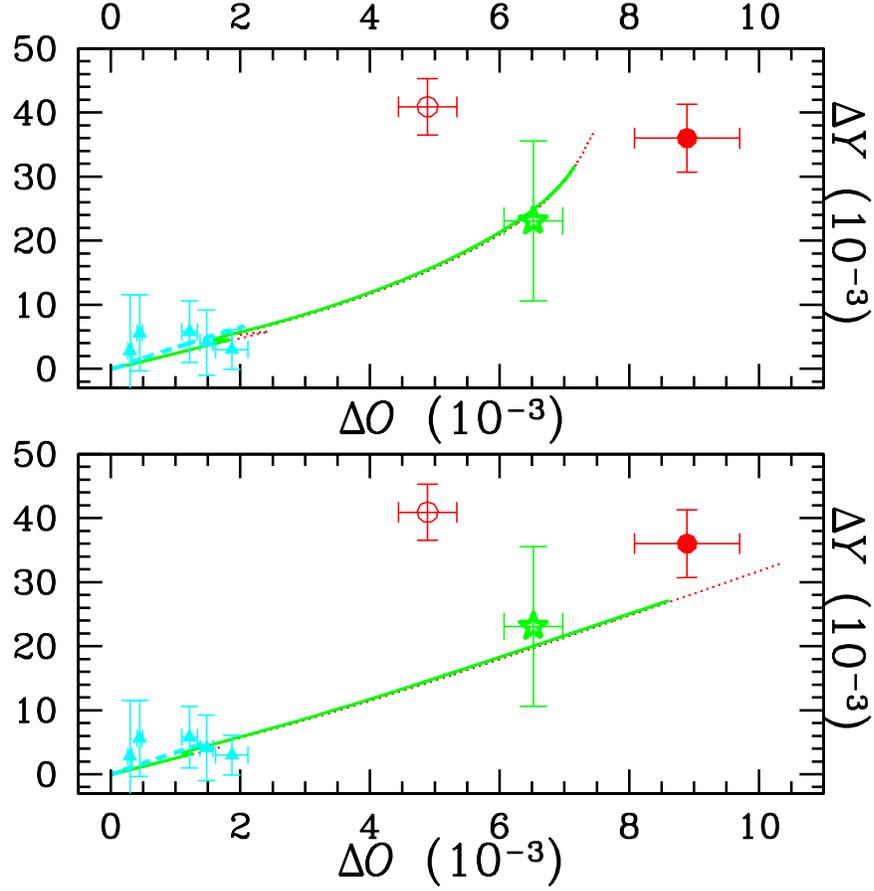}
\caption{Evolution of $\Delta Y$ vs $\Delta O$. The HWY model is shown
in the {\it upper panel} and the LWY model in the {\it lower panel}.
Chemical evolution tracks from 0 ($\Delta Y = \Delta O = 0 $) to 13 Gyr
(present time) for $r$ = 6.75, 8, and 17 kpc (dotted-red, solid-green,
and dashed-cyan lines, respectively).  The three tracks in each panel
partially overlap, from top to back they lie as follows: 17kpc,  8kpc,
and 6.75kpc.  The values for the H~{\sc ii} regions and the B stars
correspond to 13 Gyr. The M17 H~{\sc ii} region at $r$ = 6.75 kpc is
represented by a {\it filled circle}, for $t^2 = 0.036$ (RL) and by an
{\it open circle} for $t^2 = 0.00$ (FL). The solar vicinity B stars value
by \citet{prz08} is represented by a {\it star}. The extragalactic low
metallicity H~{\sc ii} regions are represented by {\it filled triangles}
for $t^2 \neq 0.00$ (FL) \citep{pei07}. A  $Y_p$ = 0.2477 is adopted
\citep{pei07}.} \label{fig:HeliumOHLWY} \end{figure}

In Figure 4 we present the evolution of $\Delta Y$ versus $\Delta O$.
The models presented in this figure for $r$ equal to 6.75, 8, and 17 kpc are
for the 0 to 13 Gyr range. The data should be compared with the end point
of the evolution of the corresponding evolutionary track.  In Figures 4,
5 and 7 the evolutionary tracks and the related observational data have
the same color.

 Since there are no good abundance determinations for the outer
parts of the Galactic disk, we test our models with data of irregular
galaxies. Therefore, in Figure 4 we show the $\Delta Y$ and $\Delta O$
values of the metal poor extragalactic H~{\sc ii} regions determined
by \citet{pei07}.  Specifically, we compare our 17kpc track with NGC
2366, because it contains one of the brightest H~{\sc ii} regions of the
Galactic vicinity.  The gaseous $O$ abundances of the metal poor H~{\sc
ii} regions were increased by 0.10 dex to include the fraction of O atoms
embedded in dust grains inside the H~{\sc ii} regions \citep{pea10}.
 From that figure it can be noticed that the predicted chemical
evolution of the Galaxy for large galactocentric radii ($r>17$ kpc) at $t = 13$
Gyr behaves like irregular galaxies at present-time.

In Figure 4 we also present the abundances of B stars of the solar
vicinity derived by \citet{prz08}.  We have adopted the Orion $X$ value
shown in \citet{car06} for converting Xi/H  (the abundance ratio by
number of any i element) of B stars to $X$i by mass.  The $\Delta Y$
abundance derived from B stars is in good agreement with the HWY and LWY
models for $r$ = 8 kpc. On the other hand the $\Delta O$ value is about
1.5 $\sigma$ and about 4.5 $\sigma$ smaller than the values predicted
by the HWY and LWY models. The O/H values derived by \citet{sim10} for
the B stars of the Orion star forming region are in good agreement with
the \citet{prz08} determinations.

We decided to include as observational constraints the O and He abundances
of the H~{\sc ii} region M17. This object has the best He abundance
determination available because its degree of ionization is very high
and the correction for neutral helium, that is always indirect, is the
smallest for the well observed Galactic H~{\sc ii} regions (see Paper I).
 We show in Figure 4 two sets of $\Delta O$ values, one derived from O
recombination lines, and the other derived from the O forbidden lines 
under the assumption that $t^2$ = 0.00.

The point with the highest $\Delta Y$ and $\Delta O$ values of the HWY
model at 6.75 kpc presented in Figure 4 (the point at $t =$ 13 Gyr),
should be compared with the two M17 values. The $\Delta Y$ model value is
in good agreement with the two M17 values. The $\Delta O$ value for $t^2$
= 0.00 is more than 5$\sigma$ smaller than the model prediction, while
the $t^2\ne0.00$ point is less than 2$\sigma$ higher than the $\Delta O$
model prediction. For the LWY model the predicted $\Delta Y$ value is in
good agreement with the two M17 values; but the $\Delta O$ value of the
$t^2$ = 0.00 point is $12\sigma$ smaller than the model prediction, while
the $\Delta O$ value of the $t^2\ne0.00$ point is less than 2$\sigma$
smaller than the model prediction. Based on this comparison we conclude
that it is possible to get an excellent agreement with the $t^2\ne0.00$
$\Delta O$ value using a model with intermediate yields between the HWY
and the LWY models, but that it is not possible to find a reasonable
model to fit the $\Delta O$ value derived assuming $t^2$ = 0.00.

Since the HWY and the LWY models fit the protosolar values and produce
a reasonable fit to the observed M17 $\Delta O$ and $\Delta Y$ values, we
conclude that the protosolar values in the context of Galactic chemical
evolution provide a strong consistency check to the O recombination
abundances derived from H~{\sc ii} regions and are in disagreement with
the O abundances derived from forbidden lines under the assumption
that $t^2$ = 0.00. Additional support for this result is provided by
\citet{sim11} who based on a study of 13 B-type stars of the Orion star
forming region OB1 find that the stellar O/H abundances agree much better
with the Orion H~{\sc ii} region abundance derived from recombination
lines than with the one derived from collisionally excited lines.


\subsection{$\Delta C$ vs $\Delta O$ evolution compared with H~{\sc ii}
regions, and young B, F, and G stars of the solar vicinity}

\begin{figure}[!t] \includegraphics[width=\columnwidth]{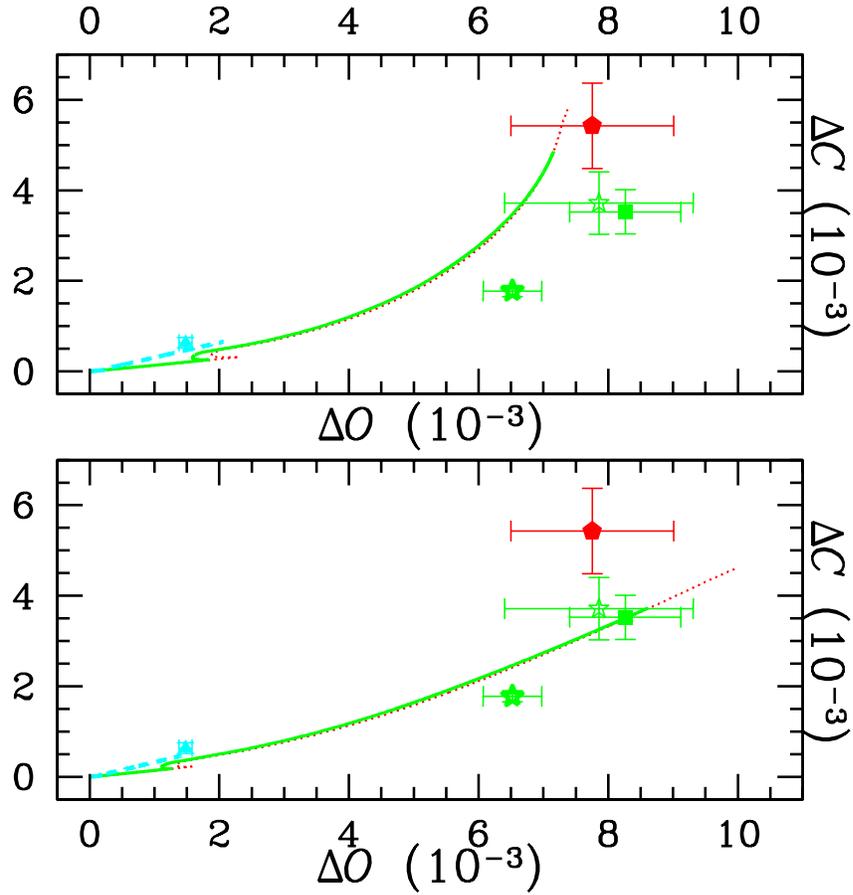}
\caption{$\Delta C$ vs $\Delta O$ diagram. The dotted red line represents
the chemical evolution track from 0 ($\Delta C = \Delta O = 0 $) to 13
Gyr (present time) for $r=$ 7 kpc, the other lines as in Figure 4. A
description of the symbols follows.  {\it Filled pentagon:} Average
values of the M17 and M20 H~{\sc ii} regions at $r=$ 6.75 and 7.19 kpc,
respectively.  {\it Empty star}: Average values of young F-G dwarf stars
of the solar vicinity, $r \sim 8 $ kpc, from \citet{ben06} (see text).
{\it Filled square}: Average values of NGC 3576 and Orion H~{\sc ii}
regions at  $r=$ 7.46 and 8.40 kpc.  {\it Filled star}: Average value
of B stars by \citet{prz08}.  {\it Filled triangle}: NGC 2363 H~{\sc
ii} region in a very  metal-poor irregular galaxy (see text).  }
\label{fig:CarbonOHLWY} \end{figure}

In what follows we will discuss the $\Delta C$ and $\Delta O$
observational values adopted from the literature and we will compare
them with the HWY and LWY model predictions.  In Figure 5 we show the
0-13Gyr evolution of the $\Delta C$-$\Delta O$  relation for $r$ =
7, 8, and 17 kpc.

The O/H ratio of the LWY model for $r$ = 17 kpc and $t$ = 13 Gyr,
corresponds to the O/H value of NGC 2363, the best extragalactic
metal-poor H~{\sc ii} region for our purpose.  This model is in good
agreement with the C/O observed ratio. Similarly the O/H ratio of the
HWY model for $r$ = 17 kpc and $t$ = 13 Gyr, corresponds to the O/H value
of NGC 2363, this model is also in good agreement with the C/O observed
ratio. The $\Delta O$ gas and dust values for NGC 2363 are those presented
in Figure 4, while the $\Delta C$ value includes  gas \citep{est09} and dust components. 
We have considered the $Y$ and $O/Z$
values from \citet{pei07} to calculate $X$ for NGC 2363 and to change
the abundances by number to abundances by mass.

For  $r \sim 8$ kpc and $t$ = 13 Gyr there are several reliable C and O
abundance determinations of the solar vicinity. To compare our models
with observations we chose: a) the average C/H and O/H values of NGC
3576 and Orion H~{\sc ii} regions at $r$= 7.46 and 8.40 kpc and the
correction for the fractions of C and O embedded in dust grains, b) the
young F and G dwarf stars studied by \citet{ben06}, and c) the average
C and O values of B stars by \citet{prz08}.

In Figure 5 we compare our models with the average $\Delta C$ and $\Delta
O$ values of NGC 3576 and Orion, and find a reasonable agreement. To
convert the Xi/H values by number to $X$i by mass we adopted the
$X$(Orion) value shown by \citet{car06}.

\citet{ben06} studied 51 F and G dwarf stars of the solar vicinity,
35 belonging to the thin disk and 16 to the thick disk. To compare
the abundances predicted by our models with those of the youngest stars
presumably recently formed, it is necessary to select the youngest subset
of the thin disk stars. The metal richest stars are expected to be the
youngest ones. We made three subsets of the metal richest stars of the
thin disk containing 4, 8 and 16 objects respectively, and obtained  
12+log(O/H) average values of 8.87, 8.84, and 8.83 respectively, 
and 12+log(C/H) average values of 8.63, 8.64, and 8.59 respectively.  In Figure
5 we present the subset of 8 stars as representative of the present
day ISM, where we have adopted the $X$(Orion) by mass \citep{car06}
for converting Xi/H to $X$i by mass. The agreement is very good. As we
saw above the other two subsets of F and G stars produce similar C/H
and O/H values also in good agreement with our models.

To convert the C and O abundances by number to abundances by mass of
the B stars by \citet{prz08} we have taken the Orion $X$ value shown
in \citet{car06}. From Figure 5 it can be seen that the $\Delta C$
value derived for the B stars is many sigma smaller than predicted by
the HWY and the LWY models. Moreover it is also considerably smaller
then the values derived from the F-G dwarf stars, and the average
of NGC 3576 and the Orion nebula. Maybe part of the reason for the
smaller C abundance in the B stars can be due to rotational mixing
\citep[e.g.][]{mey00,fie08}. The B stars $\Delta O$ value is about
1.5$\sigma$ and about 4.5$\sigma$ smaller than the predicted HWY and LWY
model values respectively. Furthermore the B stars $\Delta O$ value is
in fair agreement with the F-G dwarf value and about 1.5$\sigma$ smaller
than the the average value of NGC 3576 and Orion. These differences
between the B stars and the other objects of the solar vicinity should
be studied further.

For  $r$ = 7 kpc and $t$ = 13 Gyr we took the average of the abundances
of the M17 and M20 H~{\sc ii} regions as the observational constraint,
the C/H and  O/H gaseous values were obtained from \citet{gar07} and
Paper I using recombination lines of C and O. We have adopted the M17 $X$
value, obtained in Paper I, to convert the abundances by number of M17
and M20 to abundances by mass. Unfortunately there are no observational
data representative of the chemical past at $r$ = 7 kpc.

The HWY model adjusts very well the current  $C$ and $O$ values for $r
= 7$ kpc and marginally the present-day $O$ for $r= 8$ kpc, but does
not explain the current  $C$ value for $r= 8$ kpc.  Alternatively the
LWY model adjusts quite well both the $C$ and $O$ values for the solar
vicinity, but predicts a higher $O$ abundance for $r = 7$ kpc.

\begin{figure}[!t] \includegraphics[width=\columnwidth]{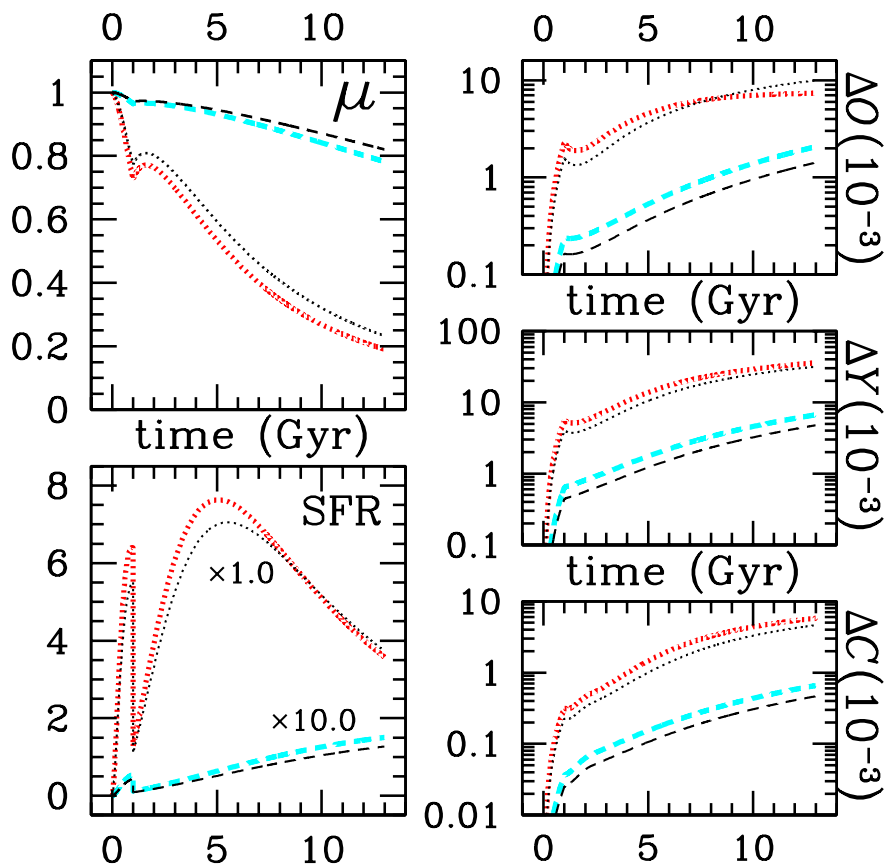}
\caption{ Evolution of $M_{gas}/M_{tot} = \mu$, star formation rate,
$\Delta O$, $\Delta Y$, and $\Delta C$ for  $r = 7$ kpc (dotted
lines) and  $r = 17$ kpc (dashed lines), assuming  HWY (thick tracks) or 
LWY (thin tracks).  Star formation history is in \msun pc$^{-2}$ Gyr$^{-1}$ units
and the SFR for $r = 17$ kpc has been multiplied by 10.
 } \label{fig:muaoayac} \end{figure}

In Figures 4 and 5 the evolutionary tracks partially overlap, 
for a given $\Delta O$ value the $\Delta Y$ and $\Delta C$ values of the two different tracks 
are almost the same, but the time is very different. In
Figure 6 we present the evolution of $\Delta O$, $\Delta Y$, and $\Delta
C$ as a function of time to appreciate the differences between the 7kpc
and the 17kpc tracks. It is important to
present the behavior of $\mu=M_{gas}/M_{tot}$ because it drives the $O$ evolution. 
For example the 17kpc track reaches a $\Delta O$ value of about 2 $\times
10^{-3}$ at 13 Gyr, while the 7kpc track reaches this value at less
than 1 Gyr, where the same $\Delta O$ value for both tracks occurs at a
similar  $\mu$ value.  The small differences in $\Delta O$, $\Delta Y$,
and $\Delta C$ at a given $\mu$ value come from: a) the delay in the
C and He enrichment of the ISM due to LIMS, b) the shape of the SFR,
and c) the differences between the HWY and LWY. In Figure 6 we present the
SFR behavior where it can be seen that for $r=7$ kpc the SFR is not only 
different in shape to the one at $r=17$ kpc but it is from one to two orders
of magnitude higher.  

From the HWY model it has been found that about half of $\Delta
Y$ and $\Delta C$ have been formed by LIMS and half by MS, \citep[see
Figure 3 of][and Figure 6 of Paper I]{car05}, while most of the
$\Delta O$ has been formed by MS.  LIMS produce a very small amount of O
according to yields by different authors \citep[see Figure 7 of][]{kar07},
in this paper we are using the yields by Marigo and collaborators.  
For the LWY model the fractions of  $\Delta Y$ and $\Delta C$ due to MS
are similar but smaller than for the HWY model. The similar fraction of
C and He produced by MS and LIMS is also responsible for the behavior
in Figures 4 and 5 where it can be seen that the $\Delta Y/\Delta O$
and the $\Delta C/\Delta O$ relations are similar for the HWY model and
for the LWY model for $\Delta O < 5\times 10^{-3}$.


\section{Intermediate Wind Yields} \label{sec:intermediate}

From Figures $1-5$ we conclude that the HWY and LWY models are in very
good agreement with some of the data, but for other data the agreement is
only fair. To improve the agreement with the observations we suggest the
use of intermediate wind yields (IWY) for the computation of the chemical
evolution models. We define the IWY as the average of HWY and LWY.

\begin{figure}[!t] \includegraphics[width=\columnwidth]{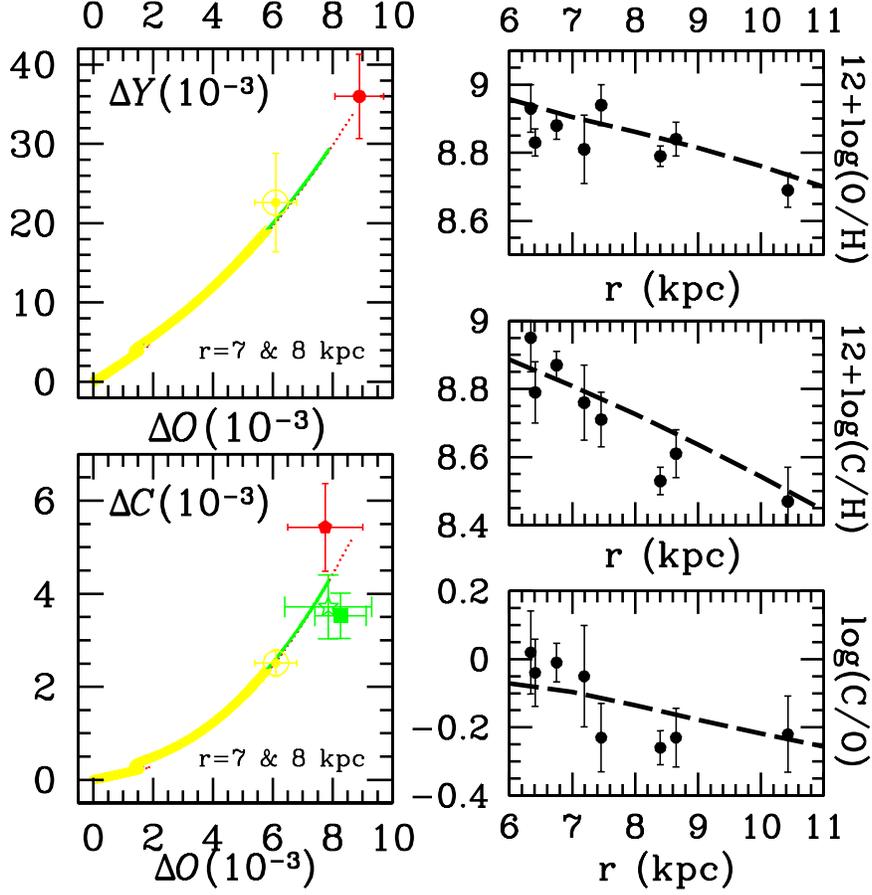}
\caption{ Chemical evolution model that assumes IWY=(HWY+LWY)/2.0.
The left panels show the 0-13Gyr evolution of $\Delta Y$ vs $\Delta O$
and of $\Delta C$ vs $\Delta O$ for $r = 7$ kpc (dotted red lines)
and $r = 8$ kpc (thin solid green line), data as Figs. 4 and 5; the
thick yellow lines show the evolution from 0 to 8.5 Gyr for $r$ = 8 kpc,
data as in Fig. 3. The right panels show the present-day ISM abundance
ratios as a function of galactocentric distance, data as in Fig. 1.}
\label{fig:GradientIntermediate} \end{figure}

In Figure 7 we present the chemical abundance predictions derived
from the IWY model for the most critical observations. In particular,
it is remarkable the fit of the model with the observed values for: a)
the present-day ($t$ = 13 Gyr) C/O gradient in the $ 6 < r$(kpc)$ < 11$
range with the gradient derived from Galactic H~{\sc ii} regions, b)
the $t$ = 13 Gyr $\Delta Y$, and $\Delta O$ values for $r=7$ kpc with
those of M17, c) the $t$ = 13 Gyr $\Delta C$, and $\Delta O$ values
for $r=7$ kpc with the average of the M17 and M20 values, d) the $t$ =
13 Gyr $\Delta C$, and $\Delta O$ values for $r=8$ kpc with the average
values for NGC 3576, and Orion, and for young F and G stars of the solar
vicinity, and e) the $t=8.5$ Gyr $\Delta Y$, $\Delta C$, and $\Delta O$
values for $r=8$ kpc  with the protosolar values.

The fits presented in Figure 7 imply that the effects of migration have
not been very important in the history of the chemical evolution of the
Galaxy.  A similar conclusion on the migration effects for the solar vicinity has been found by
\citet{nav11} based on the kinematic properties and the metallicity of thin disk stars.


\section{Possible Implications for Other Systems} \label{sec:other}

In is important to study how general is the chemical evolution model
derived for the disk of the Galaxy and to explore its relationship with
other spiral galaxies and the transition region between the disk and
the bulge of the Galaxy. In what follows we present a preliminary
discussion of these two topics.

\subsection{Other spiral galaxies}

\begin{table}[!t]
\centering
\setlength{\tabnotewidth}{0.8\columnwidth}
\tablecols{7}
\caption{Extragalactic H~{\sc ii} Regions}
\label{tta:extrahii}
\small
\begin{tabular}{lcccccc}
\toprule
Host Galaxy & Type of Galaxy & Object & $r$ (kpc) & 12+log(O/H)\tabnotemark{a} & 12+log(C/H)\tabnotemark{a} & log(C/O)\tabnotemark{a} \\
\midrule
M 101 & ScdI      & H1013    &  5.50 & 8.85 $\pm$ 0.09 & 8.67 $\pm$ 0.12 & $-0.08 \pm$ 0.15 \\
      &           & NGC 5461 &  9.84 & 8.61 $\pm$ 0.06 & 8.30 $\pm$ 0.20 & $-0.21 \pm$ 0.22 \\
      &           & NGC 5447 & 16.21 & 8.64 $\pm$ 0.06 & 8.20 $\pm$ 0.12 & $-0.34 \pm$ 0.14 \\
      &           & NGC 5471 & 23.45 & 8.35 $\pm$ 0.15 & 7.79 $\pm$ 0.19 & $-0.46 \pm$ 0.24 \\
M 33  & ScdII-III & NGC 595  &  2.87 & 8.81 $\pm$ 0.05 & 8.63 $\pm$ 0.12 & $-0.18 \pm$ 0.13 \\
      &           & NGC 604  &  4.11 & 8.72 $\pm$ 0.03 & 8.40 $\pm$ 0.11 & $-0.22 \pm$ 0.12 \\
M 31  & SbI-II    & K932     &  16.0 & 8.74 $\pm$ 0.03 & 8.49 $\pm$ 0.13 & $-0.18 \pm$ 0.14 \\
NGC 2403 & ScdIII & VS44     &  2.77 & 8.73 $\pm$ 0.04 & 8.32 $\pm$ 0.18 & $-0.31 \pm$ 0.19 \\
\bottomrule
\tabnotetext{a}{Gaseous value plus dust correction.}
\end{tabular}
\end{table}

\begin{figure}[!t] \includegraphics[width=\columnwidth]{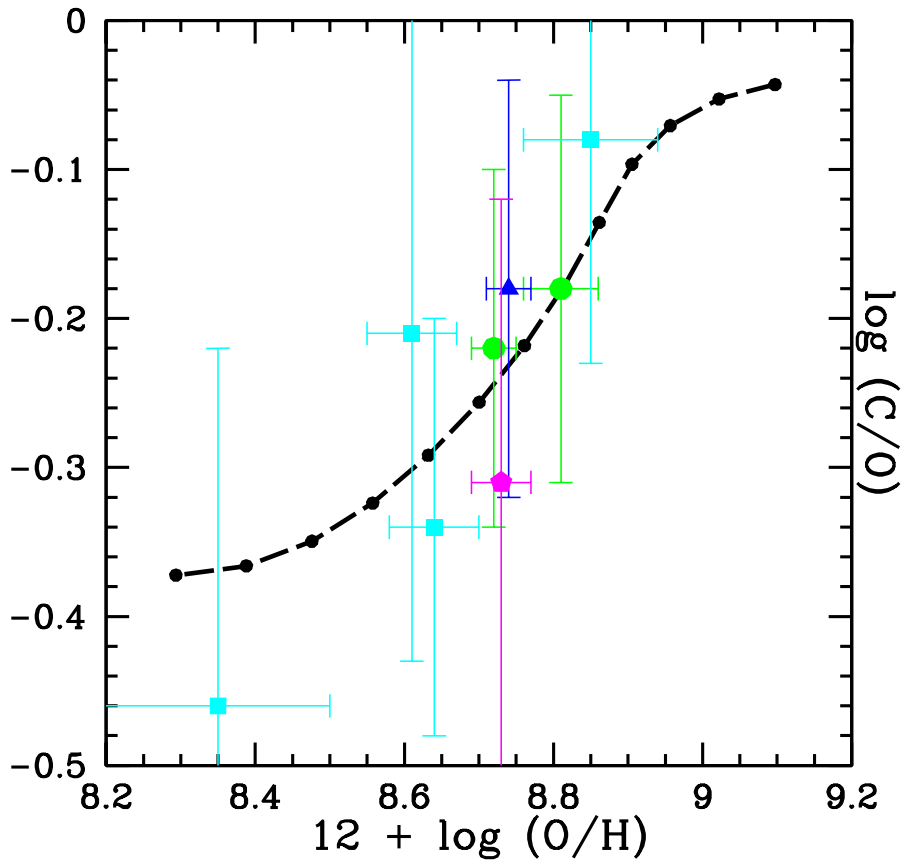}
\caption{ C/O-O/H values at 13 Gyr predicted by IWY model (long-dashed
lines).
 {\it Small black circles}: model predictions at different Galactocentric
distances, from 4 kpc (right) to 16 kpc (left). Observed abundances
ratios of extragalactic H~{\sc ii} regions in spiral galaxies: {\it
cyan squares}: M101, {\it big green circles}: M33, {\it blue triangle}:
M31, and {\it magenta pentagon}: NGC 2403.  } \label{fig:Extragalactic}
\end{figure}

Since the IWY model successfully reproduces the He, C, and O
abundances in the Galactic disk and in the solar vicinity, we want to
extend our study to other spiral galaxies with abundances determined by
recombination lines in H~{\sc ii} regions.

In Table 1  we collect the gaseous abundances of 8 extragalactic H~{\sc
ii} regions that belong to four spiral galaxies: M31, M33, M101, and NGC 2403
\citep{est02,est09}, where we have included  dust corrections identical
to those considered for  Galactic H~{\sc ii} regions.  Consequently, these
data are consistent with the Galactic data used in our previous figures
(e.g. Figure 7). In Table 1, we show the name and type of the host galaxy, and
the name and the galactocentric distance of each extragalactic H~{\sc ii} region.

In Figure 8 we show the present-day C/O and O/H values for $r = 4, 5 ,
..., 15, 16$ kpc (from right to left) predicted by the IWY model of the
Galactic disk and the  extragalactic H~{\sc ii} regions values
shown in Table 1.  

We notice that in the C/O-O/H relation presented in Figure 8: a) the
spiral galaxy disks have a similar behavior to that of the Milky Way
disk, b) all the extragalactic H~{\sc ii} regions of the sample are well
reproduced by IWY models, and c) we see the effect of the $Z$ dependent
yields for 12+log(O/H) $>$ 8.4. More determinations of the O/H and C/H ratios 
in extragalactic H~{\sc{ii}} regions with 12+log(O/H) values $< 8.4$ and $> 8.8$ 
are needed to test the yields further. 

 Since the general C/O-O/H trend of the extragalactic H~{\sc ii}
regions can be addressed using the results of the GCE model, we can
suggest that the four extragalactic spiral galaxies have a similar IMF,
no selective outflows, and probably an inside-out formation scenario
like our galaxy.  

It should be clear that Figure 8 does not correspond to models of the
four spiral galaxies, and that the $r$ values correspond to our Galaxy
and not to the other galaxies.

In Table 2 we present some galactic information of M33, M101, and
the Milky Way, the only galaxies with determined chemical gradients based
on recombination lines. Specifically, we show the galactic photometric
radius to 25 mag per square second ($R_0$) and the chemical gradients
normalized to $R_0$ corrected by dust depletion \citep{est02, est05,
est09}. The MW $R_0$ was taken from \citet{dev78}.

In Figure 9 we show the O/H, C/O, and C/O values as a function of $r$
normalized to $R_0$ for the 8 extragalactic and the 8 Galactic H~{\sc ii}
regions presented in Figs. 8 and  7, respectively.  Moreover we represent
the chemical gradients shown in Table 2.

From Figure 9, it can be noted that: 
i) the O/H slope for the three galaxies is the same within the errors, 
which suggests that for a given galaxy the O/H gradient normalized to $R_0$ 
is a more meaningful representation than the standard gradient;  
ii) the O/H gradients extend from $r/R_0 \sim 0.2$ to $r/R_0 \sim 1.0$; 
iii) the O/H ratio at a given $r/R_0$ differs by a constant among M33, M101 and the MW;
and iv) for 12+log(O/H)$\sim$8.8 the slopes of the C/H and C/O gradients become steeper.

The  parallel behavior of the O/H gradients when they are
plotted relative to $r/R_0$ is remarkable, suggesting that the main mechanisms
involved in the formation and evolution of spiral galaxies are similar. 

The slope of the O/H gradients is similar, but the O/H  absolute
value at a given $r/R_0$ is different for each galaxy. In particular the MW shows a 
ratio 0.24 dex higher O/H than the average of M33 and M101
at a given $r/R_0$. Considering that the MW is an Sbc galaxy and M33 and
M101 are Scd galaxies, this result probably implies that the earlier the
type of a galaxy the higher the O/H value at a given $r/R_0$, 
this result might be due to the fact that the
earlier the bulk of star formation the fainter the stellar
luminosity at a given O/H value and consequently the higher
the O/H value at a given $R_0$ value. 
Considering that the MW is an Sbc galaxy and M33 and M101 are Scd galaxies, 
this result might be of a general nature and suggests that the
earlier the type of a galaxy the higher the O/H value at a
given $r/R_0$. 
More high accuracy determinations of O gradients
in spiral galaxies are needed to test this result.

\begin{table}[!t]
\centering
\setlength{\tabnotewidth}{0.8\columnwidth}
\tablecols{5}
\caption{Abundance gradients normalized to $R_0$\tabnotemark{a}}
\label{tta:extragrad}
\begin{tabular}{lcccc}
\toprule
Galaxy & $R_0$(kpc)\tabnotemark{b} & 12+log(O/H) & 12+log(C/H) & log(C/O) \\
\midrule
M101 & 28.95 & $-0.492\times$(r/$R_0$)+8.84 & $-1.32\times$(r/$R_0$)+8.99 & $-0.65\times$(r/$R_0$)+0.03 \\
M33  &  6.83 & $-0.492\times$(r/$R_0$)+9.00 & $-0.72\times$(r/$R_0$)+8.93 & $-0.22\times$(r/$R_0$)$-0.10$ \\
MW   & 11.25 & $-0.495\times$(r/$R_0$)+9.16 & $-1.16\times$(r/$R_0$)+9.50 & $-0.65\times$(r/$R_0$)+0.34 \\
\bottomrule
\tabnotetext{a}{$R_0=$ galactic photometric radius to 25 mag per square second.}
\end{tabular}
\end{table}

\begin{figure}[!t] \includegraphics[width=\columnwidth]{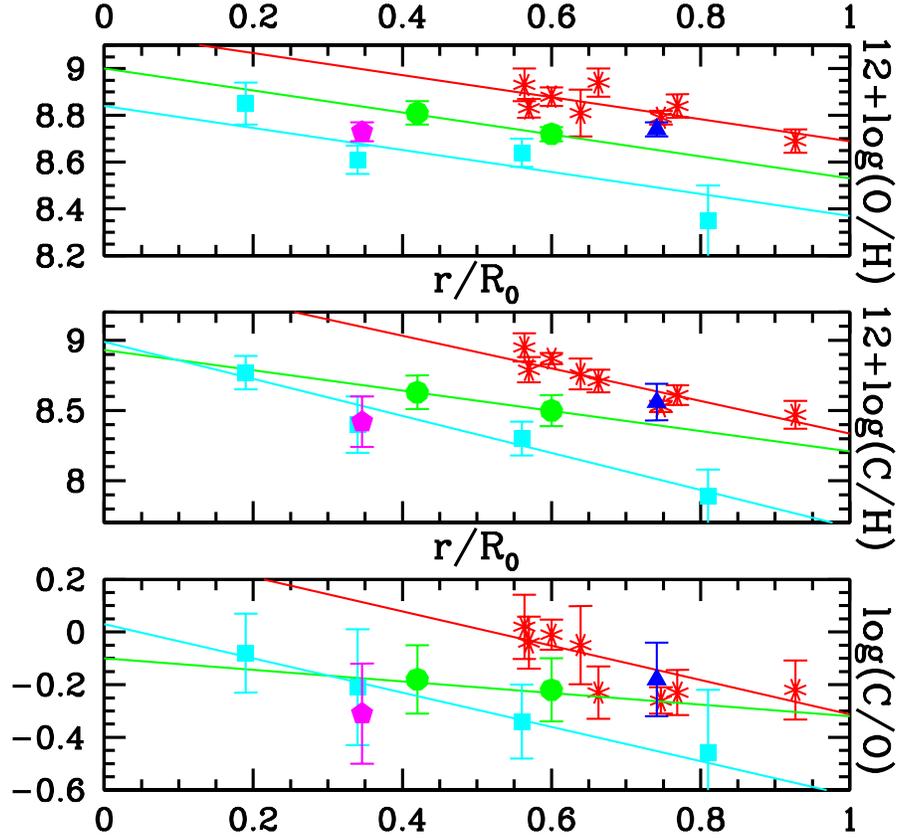}
\caption{ Chemical abundance ratios of  H~{\sc ii} regions normalized to
galactic photometric radius ($R_0$) for each galaxy (see Tables 1 and 2).
Milky Way: {\it red asterisks} (see Fig. 7).  Others spiral galaxies:
symbols as Fig. 8.  Lines represent the galactic disk gradients of M101,
M33, and the Milky Way.  } \label{fig:Extragrad} \end{figure}

From Figures 8 and 9 it follows that we need: i) to increase the
$r/R_0$ coverage, mainly in the shorter galactocentric distance,
with  abundance determinations of high accuracy, ii) to increase the sample
by including spiral galaxies of different types, and iii) to make specific
models for each galaxy.  All in order to sort out possible differences in
the galactic formation and evolution relative to that of the Galaxy and
to understand the peculiar behavior of metal-rich stars and H~{\sc ii}
regions, see \citet{car08b}, and to be able to test stellar yields for $Z>\Zsun$.

The present paper only includes specific models for our galaxy, we
present a preliminary discussion on the probable relevance of  our
results to the study of other spiral galaxies. It is beyond the scope
of this paper to produce models for other galaxies and therefore to
present a detailed comparison with the results derived by other authors,
e.g. \citet{chi03b,mol05,ren05,yin09,mar10,mag10}, using different
sets of data.  In future papers we plan to present detailed models for
M31 \citep{men11} and M33 (Robles-Valdez, Carigi, \& Bruzual 2011 
in preparation). 


\subsection{The transition region between the inner disk and the Galactic bulge}

\begin{figure}[!t] \includegraphics[width=\columnwidth]{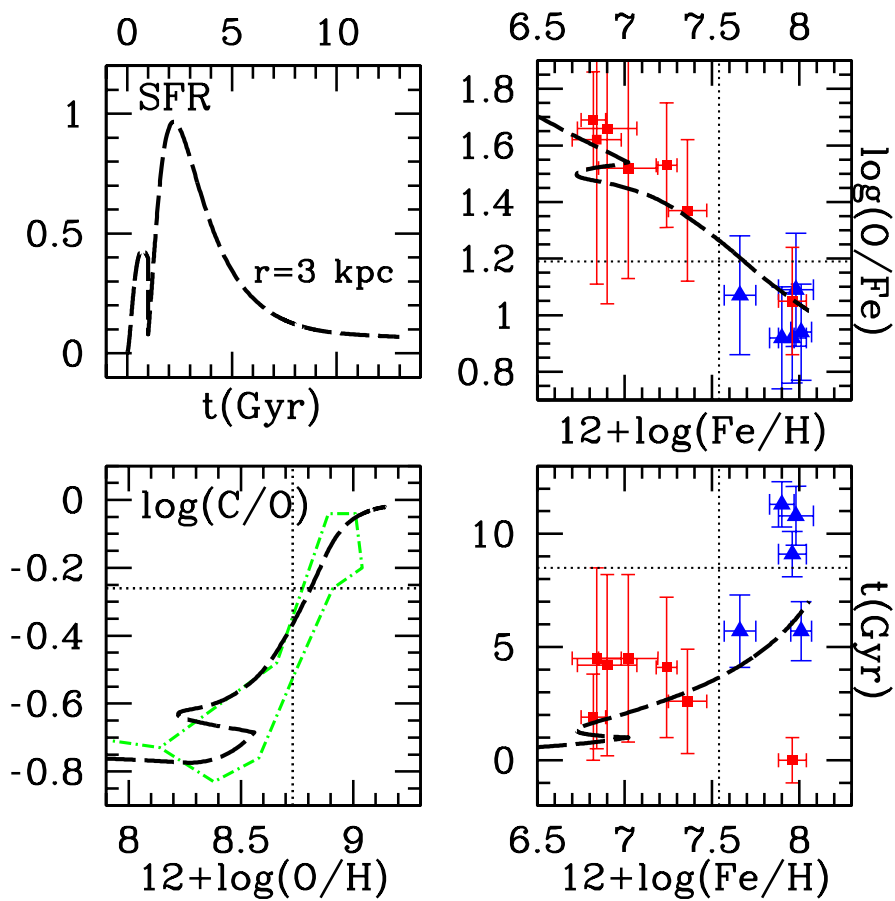}
\caption{ Evolution for $r = 3$ kpc predicted by the IWY model (long-dashed
black lines).  Star formation history normalized to 70 \msun pc$^{-2}$
Gyr$^{-1}$.  C/O-O/H, O/Fe-Fe/H, and time-Fe/H relations from 0 to
7 Gyr.  {\it Area enclosed by dot-short-dashed-green lines}: Galactic
bulge red giant stars presented by Cescutti et al. (2009).  {\it Filled
symbols}: microlensed dwarf and subgiant stars in the Galactic bulge
from \citet{ben10a}, {\it red squares}: old stars with mean age of
9.8 $\pm$ 4.0 Gyr, {\it blue triangles}: young stars with mean age
of 4.5 $\pm$ 1.6 Gyr.  Ages were normalized to the age of the models
(13 Gyr.)  {\it Dotted lines}: protosolar values from \citet{asp09}.}
\label{fig:Bulge} \end{figure}

A fraction of the stars observed in the direction of the Galactic center might
have been formed in the true bulge and another fraction in the inner Galactic disk. 
Therefore we considered interesting to compare our IWY model for the innermost 
part of the disc  with the abundances of these stars.

Since the inside-out model adopted as the Galactic formation scenario
in this paper diverges for $r=2$ kpc (see Section 2), we obtained the
chemical evolution of the  innermost ring of the Galactic disk that our
GCE model is able to consider, $r=3 \pm 0.5$ kpc (with a 1 kpc width).
The Galaxy at this radius formed efficiently: during the first Gyr the
halo formed with a collapse time-scale of 0.5 Gyr and then the disk formed
with a formation time scale of 1.0 Gyr.  At the end of the evolution
($t$ = 13 Gyr) the gaseous mass amounts to 7\% of the baryonic mass
inside the $r=3 \pm 0.5$ kpc ring.

In Figure 10 we show the C/O-O/H, O/Fe-Fe/H, and Fe/H evolution obtained
with the IWY model between 0 and 7 Gyr.  We include the evolution from
0 to 7 Gyr and exclude the 7-13 Gyr range because at 7 Gyr the model
reaches the highest Fe/H value of the stellar data. Moreover, in the excluded
range the SFR is very low and the probability to form stars younger than
6 Gyr (in the 7 to 13 Gyr range) is negligible.

In Figure 10 (C/O-O/H panel) we show also the area covered by the data
of \citet{ces09}.  These authors collected [C/O] and [O/H] ratios for the
Galactic bulge red giants determined by \citet{ful07} and \citet{mel08},
and used the observed data to obtain the initial stellar C/H and O/H
values relative to the solar abundances due to \citet{asp05}.  In order
to convert the Cescutti et al. data to abundances by number we used,
only in this figure, the solar abundances by \citet{asp05}. For the
other figures, we assumed the protosolar abundances by \citet{asp09}.

A description of the model in the C/O-O/H panel of Figure 10 follows.
During the halo formation, the IWY model predict an increase of
12+log(O/H) to 8.6 dex and  an increase of 12+log(C/O) to -0.7 dex .  
After the halo formation
finishes, at $t$ = 1 Gyr, the O/H ratio decreases due to the dilution
produced by the enormous amount of primordial gas that is accreted to
form the disk.  Later on this accretion causes a rapid SFR increase,
producing an O/H increase. The C/O rise from -0.7 dex to -0.5 dex is due mainly to LIMS,
while the C/O rise from -0.5 dex to 0.0 dex is due to both, massive stars and LIMS.

In Figure 10 (O/Fe-Fe/H and time-Fe/H panels), we included the chemical
abundances determined by \citet{ben10a} in microlensed dwarf and subgiant
stars of the Galactic bulge.  In order to convert the data by Bensby et
al. to abundances by number, we used their solar abundances.  The stellar
ages were normalized to the age of the model ($t=13.0$ Gyr).

Our model at $r$ = 3 kpc successfully
reproduces the O/Fe-Fe/H obtained by \citet{ben10a}, but the time-Fe/H
relation is only partially reproduced.  Our model explains all
old-metal-poor stars and two young stars, but not the oldest metal-rich
star in their sample (see the square at 12+log(Fe/H)$\sim$ 7.9) which
behaves like a common bulge star, see \citet{bal07}.  The youngest and
metal-rich stars would be disk stars that formed at $r<3$ kpc. 

Since Fe is produced by massive stars and binary systems of LIMS, to a
first approximation Fe behaves like C, that is produced by MS and single
LIMS, see for example \citet{ake04}, the C/O-O/H discussion presented
above can be used to explain the O/Fe-Fe/H and time-Fe/H panels of
Figure 10.

\citet{ben10b} have studied red giant stars in the inner Galactic disc
and find that the abundance trends of the inner disc agree very well
with those of the nearby thick disc, and also with those of the Galactic
bulge.  Based on the stellar results  of the thick and thin disks of
the solar vicinity, of the inner Galactic disc, and of the Galactic
bulge \citep{ben06,ben10a,ben10b}, they suggest that the bulge and the
disk have had similar chemical histories.  Moreover \citet{alv10} also
suggest that the bulge and local thick disk stars experienced similar
formation timescales, star formation rates and initial mass functions.

\citet{ces09} present a Galactic bulge model, computed by \citet{bal07},
that, in order to reproduce the total stellar mass, the iron distribution
function, and the $\alpha$/Fe-Fe/H relations (all constraints obtained
from giant stars) they assume:  a formation time scale of $\sim$ 0.1 Gyr,
as well as a star formation efficiency one order of magnitude higher
and an IMF flatter for $m>1$~\msun~than those considered in our models.
This model, with a SFR as long as 0.5 Gyr, cannot explain the ages of
the youngest bulge dwarfs found by \citet{ben10a}.

Note that, if we focus only on C/O-O/H  and O/Fe-Fe/H relations
determined in dwarf stars of the Galactic bulge we cannot distinguish
between a scenario with a very rapid infall, efficient star formation,
and a high relative formation of massive stars, that by \citet{ces09},
and another scenario with an order of magnitude less rapid infall, an
order of magnitude less efficient star formation, and a lower relative
formation of massive stars (the one presented in this paper).

Nevertheless, if we focus on the age-Fe/H relation shown by bulge dwarf
stars, our model, with an extended bulge formation time and therefore
an extended star formation history, produces a better fit to the data
than that by \citet{ces09}.

\begin{figure}[!t] \includegraphics[width=\columnwidth]{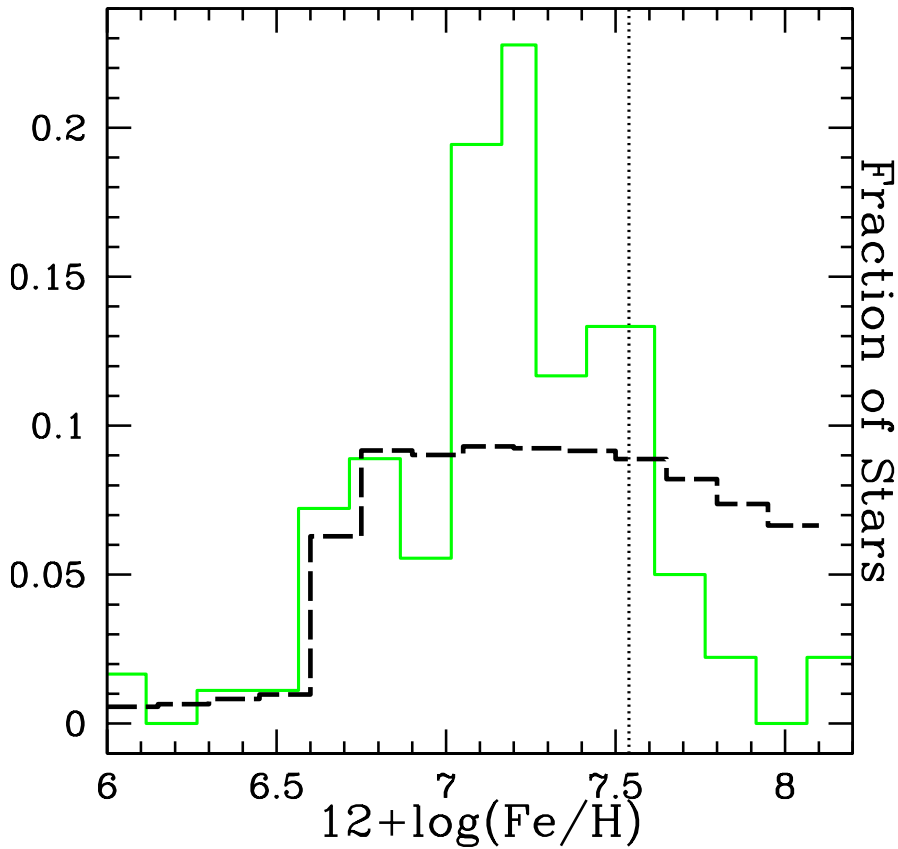}
\caption{Iron distribution function.  Predictions for $r = 3$ kpc
model that assume IWY (long-dashed black line) until 7 Gyr.  Observed
distribution for the outermost fields  (b=-12) along the bulge minor axis
by \citet{zoc08} (continuous green line).  {\it Dotted line}: protosolar
Fe/H value from \citet{asp09}.  } \label{fig:BulgeMDF} \end{figure}

 In Figure 11 we show the iron distribution function predicted by the
IWY model for $r = 3$ kpc in the Galactic disk from 0 to 7 Gyr.  Since there
are no good abundance determinations for the inner parts of the Galactic
disk, we compare our model with the 104 stars that belong to the outermost
field along the bulge minor axis at b=-12, see lower panel of Figure 14
by \citet{zoc08}.  We have used 12+log(Fe/H)$_\odot=7.50$ to convert the
[Fe/H] values of the Zoccali et al. sample to Fe/H abundances by number.
From that figure, we can say that a non-negligible fraction of the stars
in the direction of the bulge might belong to the inner disk.  Moreover,
we could comment on that bulge sample: i) for 12+log(Fe/H)$ < 7$ the
contamination of the thick disk is high, ii) for $ 7 < $ 12+log(Fe/H)$
< 8 $ the contamination of the thin disk is lower and the true bulge
stars dominate, and iii) for 12+log(Fe/H)$ > 8$ the contribution of
the metal-rich stars of the thin disk is important.  

Note that in Figure 11 we are using only the outermost 104 stars of the total
800 stars in the direction of the bulge, and that our model is for $r = 3$ kpc.
On the other hand \citet{ces11} show that their bulge model (the
same as \citet{ces09}) follows the constraint provided by the 800 stars very well
using a Salpeter IMF. The model by Cescutti and Matteucci predicts a smaller
number of stars for [Fe/H] values higher than solar, probably indicating that 
the bulge sample has been contaminated by innermost disk stars.

From the previous discussion it follows that the Galactic bulge is a
complex structure that should be studied further \citep[e.g.][]{zoc10}.


\section{Conclusions} \label{sec:conclusions}

We have made models with three different sets of yields that differ
only on their $Z$ dependence at solar metallicities for massive stars:
HWY, IWY, and LWY.  The HWY and the LWY have been used before by us,
while in this paper we introduce the IWY, that are given by (HWY +
LWY)/2. We find that the IWY galactic chemical evolution models produce
better fits to the observational data than either the HWY or the LWY
galactic chemical evolution models.

We present a Galactic chemical evolution model based on the IWY for
the disk of the Galaxy that is able to fit: a) the C/O vs O/H,
C/Fe vs Fe/H, O/Fe vs Fe/H, and Fe/H vs $t$ relations derived from halo and
disk stars of different ages in the solar vicinity, b) the O/H, C/H, and C/O
abundance gradients (slopes and absolute values) derived from Galactic
H~{\sc{ii}} regions, c) the He/H, C/H, O/H, Fe/H protosolar abundances, and
d) the He/H and O/H values of the galactic H~{\sc{ii}} region M17.

We find that in general about half of the freshly made helium is produced
by massive stars and half by LIMS, and that a similar situation prevails
for carbon, while  most of the oxygen is produced by massive stars.
The agreement of the He/O and C/O ratios between the model and the
protosolar abundances implies that the Sun formed from a well mixed ISM.

We note that the agreement of our model with the protosolar abundances
and the Sun-formation time supports the idea that the Sun originated at
a galactocentric distance similar to that of the solar vicinity.

We show that chemical evolution models for the Galactic disk are able
to reproduce the observed $\Delta Y$ and $\Delta O$ protosolar values
and the $\Delta Y$ and $\Delta O$  values derived for M17 based on H,
He and O recombination lines, but not the M17 $\Delta Y$ and $\Delta O$
values derived from $T$(4363/5007) and O collisionally excited lines under
the assumption of $t^2 = 0.00$. This result provides a consistency check
in favor of the presence of large temperature variations in H~{\sc{ii}}
regions and on the method based on the H, He, C and O recombination
lines to derive abundances in H~{\sc{ii}} regions.

We obtain that the IWY chemical evolution model of the Galactic disk
for the present time, in the galactocentric range $ 6 < r$(kpc) $<$ 11,
produces a reasonable fit to the O/H vs C/O relationship
derived from H~{\sc{ii}} regions of nearby spiral galaxies. 
The yields predict an increase of the C/O ratio with O/H starting from
12+log(O/H)$\sim$8.4 that is observed in our Galaxy and in nearby galaxies.
The O/H vs C/O relationship might
imply  that spiral galaxies have a similar IMF, no selective outflows,
and probably a formation scenario similar to that of our galaxy.

We find a remarkable parallelism of the O/H gradients
for M33, M101, and the Galaxy when they are plotted with
respect to $r/R_0$ ($R_0$ is the galactic photometric radius to 25 mag per 
square second), suggesting some common mechanisms
in the formation and evolution of spiral galaxies. The
O/H ratio at a given $r/R_0$ differs by a constant among
M33, M101 and the MW.  The MW shows a 0.24 dex higher O/H
ratio than the average of M33 and M101 at a given
$r/R_0$. 

We also find that the results for our model at $r=3$ kpc can
explain: a) the C/O-O/H and O/Fe-Fe/H, relations,
and b) partially the Fe/H-time relation and the Fe distribution function
derived from stellar observations in the direction of the Galactic
bulge. We find that stars belonging to the thin and thick discs make a
significant contribution to these relations.

Future work to advance in this subject requires:
a) to advance in the study of the galactic bulge to be able to quantify
the stellar contributions due to the inner disk and the true bulge,
b) to increase the H~{\sc{ii}} regions $r/R_0$ coverage,
mainly in the shorter galactocentric distance, with high
accuracy C/H and O/H abundance determinations;
c) to increase the sample of spiral galaxies of different
types with O/H gradients of high accuracy, and 
d) to make specific models for each galaxy.  All in order to sort out
possible differences in the galactic formation and evolution
of other galaxies relative to that of the Galaxy.

\acknowledgments

We thank Jorge Garc\'{\i}a-Rojas and C\'esar Esteban for useful
discussions.  We are also grateful to the anonymous referee for a careful
reading of the manuscript and many excellent suggestions. This work was
partly supported by the CONACyT grants 60354 and 129753.



\begin{thebibliography}

\bibitem[Alves-Brito et al.(2010)]{alv10} Alves-Brito, A., Mel\'endez,
J., Asplund, M., Ram\'{\i}rez, I., \& Yong, D. 2010, \aap, 513, 35

\bibitem[Akerman et al.(2004)]{ake04} Akerman, C. J., Carigi, L., Nissen,
P. E., Pettini, M., \& Asplund, M. 2004, \aap, 414, 931

\bibitem[Anders \& Grevesse(1989)]{and89} Anders, E. \& Grevesse, N. 1989,
Geochim. Cosmochim. Acta, 53, 197

\bibitem[Asplund et al.(2005)]{asp05} Asplund, M., Grevesse, N., \&
Sauval, A. J. 2005, in Cosmic Abundances as Records of Stellar Evolution
and Nucleosynthesis, ASP Conf. Ser. Vol. 336, eds. T. G. Barnes \& F.
N. Bash, (San Francisco, CA), 25

\bibitem[Asplund et al.(2009)]{asp09} Asplund, M., Grevesse, N., Sauval,
A. J., \&  Scott, P. 2009, Ann. Rev. A. \& Ap., 47, 481

\bibitem[Ballero et al.(2007)]{bal07} Ballero, S. K., Matteucci, F.,
Origlia, L., \& Rich, R. M. 2007, \aap, 467, 123

\bibitem[Bensby et al.(2010b)]{ben10b} Bensby, T., Alves-Brito, A., Oey,
M. S., Yong, D., \& Mel\'endez, J. 2010b, \aap, 516, L13

\bibitem[Bensby \& Feltzing(2006)]{ben06} Bensby, T. \& Feltzing, S. 2006,
\mnras, 367, 1181

\bibitem[Bensby et al.(2010a)]{ben10a} Bensby, T., Feltzing, S., Johnson,
J. A., Gould, A., et al. 2010a, \aap, 512, A41

\bibitem[Bromm \& Larson(2004)]{bro04} Bromm, B. \& Larson, R. V. 2004
ARAA, 42, 79

\bibitem[Carigi(1994)]{car94} Carigi, L.  1994, \apj, 424, 18

\bibitem[Carigi(1996)]{car96} Carigi, L.  1996, RevMexAA, 32, 179

\bibitem[Carigi(2000)]{car00} Carigi, L. 2000, RevMexAA, 36, 171

\bibitem[Carigi(2008)]{car08b}  Carigi, L. 2008 in "The Metal-Rich
Universe", eds. G. Israelian and G. Meynet.  Series: Cambridge
Contemporary Astrophysics, p.415 (arXiv:astro-ph/0612049)


\bibitem[Carigi et~al.(1999)]{car99} Carigi, L., Col\'{\i}n, P., \&
Peimbert, M. 1999, \apj, 514, 787

\bibitem[Carigi et~al.(2006)]{car06} Carigi, L., Col\'{\i}n, P., \&
Peimbert, M. 2006, \apj, 644, 924

\bibitem[Carigi et~al.(1995)]{car95} Carigi, L., Col\'{\i}n, P., Peimbert,
M., \& Sarmiento, A. 1995, \apj, 445, 98

\bibitem[Carigi \& Hernandez (2008)]{car08c} Carigi, L. \& Hernandez,
X. 2008, \mnras, 90, 582

\bibitem[Carigi \& Peimbert(2008)]{car08} Carigi, L. \& Peimbert, M. 2008,
RevMexAA, 44, 311 (Paper I)

\bibitem[Carigi et~al.(2005)]{car05} Carigi, L., Peimbert, M., Esteban,
C., \& Garc\'{\i}a-Rojas, J. 2005, \apj, 623, 213

\bibitem[Cescutti et al.(2007)]{ces07} Cescutti, G., Matteucci, F.,
Francois, P., \& Chiappini, C. 2007, \aap, 462, 943

\bibitem[Cescutti et al.(2009)]{ces09} Cescutti, G., Matteucci, F.,
McWilliam, A., \& Chiappini, C. 2009, \aap, 505, 605

\bibitem[Cescutti \& Matteucci(2011)]{ces11} Cescutti, G. \& Matteucci, F.
2011, \aap, 525, A126

\bibitem[Chiappini et al.(1997)]{chi97} Chiappini, C., Matteucci, F.,
\& Gratton R., 1997, \apj, 477, 765

\bibitem[Chiappini et al.(2003a)]{chi03a} Chiappini, C., Matteucci, F.,
\&  Meynet, G. 2003a, \aap, 410, 257

\bibitem[Chiappini et al.(2003b)]{chi03b} Chiappini, C., Romano, D., \&
Matteucci, F. 2003b, \mnras, 339, 63

\bibitem[de Vaucouleurs \& Pence(1978)]{dev78} de Vaucouleurs, G.\&
Pence, W. D. 1978, AJ, 83, 1163

\bibitem[Esteban et al.(2009)]{est09} Esteban, C., Bresolin, F., Peimbert,
M., Garc\'{\i}a-Rojas, J., Peimbert, A., \& Mesa-Delgado, A. 2009, \apj,
700, 654

\bibitem[Esteban et al.(2005)]{est05} Esteban, C., Garc\'{\i}a-Rojas,
J., Peimbert, M., Peimbert, A., Ruiz, M. T., Rodr\'{\i}guez, M., \&
Carigi, L. 2005, \apj, 618, L95

\bibitem[Esteban et al.(1998)]{est98} Esteban, C., Peimbert, M.,
Torres-Peimbert, S., \& Escalante, V. 1998, \mnras, 295, 401

\bibitem[Esteban et al.(2002)]{est02} Esteban, C., Peimbert, M.,
Torres-Peimbert, S., \& Rodr\'{\i}guez, M. 2002, \apj, 581, 241

\bibitem[Fenner \& Gibson(2003)]{fen03} Fenner, Y. \& Gibson, B. K. 2003,
PASA, 20, 189

\bibitem[Fierro \& Georgiev(2008)]{fie08} Fierro, C. L. \& Georgiev,
L. 2008, RevMexAA, 44, 213

\bibitem[Fulbright et al.(2007)]{ful07} Fulbright, J. P., McWilliam,
A., \& Rich, R. M. 2007, \apj, 661, 1152

\bibitem[Garc\'{\i}a-Rojas \& Esteban(2007)]{gar07} Garc\'{\i}a-Rojas,
J. \&  Esteban, C. 2007, \apj, 670, 457


\bibitem[Greggio \& Renzini(1983)]{gre83} Greggio, L. \& Renzini, A.,
1983, \aap, 118, 217

\bibitem[Grevesse \& Noels(1993)]{gre93} Grevesse, N. \& Noels A. 1993, in
Origin and Evolution of the Elements, eds. N. Prantzos, E. Vangioni-Flam,
\& M. Cass\'e,  (Cambridge: Cambridge Univ. Press), 15

\bibitem[Grevesse & Sauval(1998)]{gre98} Grevesse,  N. \& Sauval,
A. J. 1998, Space Sci. Rev., 85, 161

\bibitem[Hirschi(2007)]{hir07} Hirschi, R. 2007, \aap, 461, 571

\bibitem[Hirschi et al.(2005)]{hir05} Hirschi, R., Meynet, G., \& Maeder,
A. 2005, \aap, 433, 1013

\bibitem[Kalberla \& Kerp(2009)]{kal09} Kalberla, P. M. W. \& Kerp,
J. 2009, Annu. Rev. Astron. Astrophys., 47, 27

\bibitem[Karakas \& Lattanzio(2007)]{kar07} Karakas, A. \& Lattanzio,
J. C. 2007 PASA, 24, 103

\bibitem[Kewley \& Ellison(2008)]{kew08} Kewley, L. J. \& Ellison,
S. L. 2008, \apj, 681, 1183

\bibitem[Kroupa et al.(1993)]{ktg93} Kroupa, P., Tout, C. A., \& Gilmore,
G. 1993, \mnras, 262, 545

\bibitem[Lodders(2003)]{lod03} Lodders, K. 2003, \apj, 591, 1220

\bibitem[Lodders et al.(2009)]{lod09} Lodders, K., Palme, H., \& Gail,
H.-P. 2009, Abundances of the elements in the Solar System, Tr\"umper, J.E. (ed.), 
The Landolt-B\"ornstein Database, Springer-Verlag Berlin Heidelberg, 4.4 


\bibitem[Maeder(1992)]{mae92} Maeder,  A. 1992, \aap, 264, 105

\bibitem[Magrini et al.(2010)]{mag10} Magrini, L., Stanghellini, L.,
Corbelli, E., Galli, D., \& Villaver, E. 2010, \aap, 512, 63

\bibitem[Marcon-Uchida et al.(2010)]{mar10} Marcon-Uchida, M. M.,
Matteucci, F.,    \& Costa, R. D. D. 2010, \aap, 520, 35

\bibitem[Marigo et al.(1996)]{mar96} Marigo, P., Bressan, A., \& Chiosi,
C. 1996, \aap, 313, 545

\bibitem[Marigo et al.(1998)]{mar98} Marigo, P., Bressan, A., \& Chiosi,
C. 1998, \aap, 331, 564

\bibitem[Matteucci(2000)]{mat00} Matteucci, F. 2000, The chemical
evolution of the galaxy, Astrophysics and space science library, v. 253.
Kluwer Academic Publishers

\bibitem[Matteucci \& Chiappini(1999)]{mat99} Matteucci, F. \& Chiappini,
C.  1999, in Chemical Evolution from Zero to High Redshift, Proceedings
of the ESO Workshop, eds. J. R. Walsh and M. R. Rosa.
 Berlin: Springer-Verlag, p. 83

\bibitem[Matteucci et al.(1989)]{mat89} Matteucci, F., Franco, J.,
Francois, \& P., Treyer, M. 1989, Rev. Mex, Astron. Astrof, 18, 145

\bibitem[Mattsson(2010)]{mats10} Mattsson, L. 2010, A\&A, 515, 68

\bibitem[Mel\'endez et al.(2008)]{mel08} Mel\'endez, J., Asplund, M.,
Alves-Brito, A., et al. 2008, \aap, 484, L21

\bibitem[Meneses-Goytia et al.(2011)]{men11}
Meneses-Goytia, S., Carigi, L., \& Garc\'{\i}a-Rojas, J. 2011,
Astrobiology, submitted

\bibitem[Mesa-Delgado et al.(2009)]{mes09} Mesa-Delgado, A., Esteban,
C., Garc\'{\i}a-Rojas, J., Luridiana, V., Bautista, M., Rodr\'{\i}guez,
M., L\'opez-Mart\'{\i}n, L., \& Peimbert, M. 2009, \mnras, 395, 855

\bibitem[Meynet \& Maeder(2000)]{mey00} Meynet, G. \& Maeder, A. 2000,
\aap, 361, 101

\bibitem[Meynet \& Maeder(2002)]{mey02} Meynet, G. \& Maeder, A. 2002,
\aap, 390, 561

\bibitem[Moll\'a \& D\'{\i}az(2005)]{mol05} Moll\'a, M. \& D\'{\i}az,
A. I. 2005, \mnras, 358, 521


\bibitem[Navarro et al.(2010)]{nav11} Navarro, J. F., Abadi, M. G.,
Venn, K. A., Freeman, K. C., \& Anguiano, B. 2010, \mnras, submitted,
(arXiv1009.0020)

\bibitem[Nordstr\"om et al.(2004)]{nor04} Nordstr\"om, B,  Mayor, M.,
Andersen, J., Holmberg, J., Pont, F., Jorgensen, B. R., Olsen, E. H.,
Udry, S., \& Mowlavi, N. 2004, \aap, 418, 989

\bibitem[Pagel(2009)]{pag09} Pagel, B. E. J. 2009, Nucleosynthesis and
Chemical Evolution of Galaxies, Cambridge University Press.

\bibitem[Peimbert \& Peimbert(2010)]{pea10} Peimbert, A. \& Peimbert,
M. 2010, \apj, 724, 791

\bibitem[Peimbert et al.(2002)]{pei02} Peimbert, A., Peimbert, M., \&
Luridiana, V. 2002, \apj, 565, 668

\bibitem[Peimbert(1967)]{pei67} Peimbert, M. 1967, \apj, 150, 825

\bibitem[Peimbert et al.(2007)]{pei07} Peimbert, M., Luridiana, V., \&
Peimbert, A. 2007, \apj, 666, 636

\bibitem[Peimbert \& Peimbert (2011)]{pei11} Peimbert, M. \& Peimbert,
A. 2011, RevMexAA, SC, in press (arXiv: 0912.3781)

\bibitem[Peimbert et al.(2010)]{pei10} Peimbert, M., Peimbert, A.,
Carigi, L., \& Luridiana, V. 2010, in Light elements in the Universe, IAU
Symposium 268, eds. C. Charbonnel, M. Tosi, F. Primas, \& C. Chiappini,
(Cambridge: Cambridge Univ. Press), p. 91

\bibitem[Portinari et al.(1998)]{por98} Portinari, L., Chiosi, C., \&
Bressan, A. 1998, \aap, 334, 505

\bibitem[Prantzos(2008)]{pra08} Prantzos, N. 2008, EAS Publications Series,
32, 311 (arXiv:0709.0833)

\bibitem[Prantzos et al.(1994)]{pra94} Prantzos, N., Vangioni-Flam, E.,
\& Chauveau, S. 1994, \aap, 285, 132

\bibitem[Przybilla et al.(2008)]{prz08} Przybilla, N., Nieva, M. F., \&
Butler, K. 2008, \apj, 688, L103

\bibitem[Renda et al.(2005)]{ren05} Renda, A., Kawata, D., Fenner, Y., \&
Gibson, B. K. 2005, \mnras, 356, 1071

\bibitem[Romano et al.(2010)]{rom10} Romano, D., Karakas, A. I., Tosi,
M., \&    Matteucci, F. 2010, \aap, 522, 32

\bibitem[Roskar et al.(2008)]{ros08} Roskar, R., Debattista, V. P.,
Stinson, G. S., Quinn, T. R.,
 Tobias Kaufmann, T., \&  Wadsley, J. 2008, \apj, 675, L65

\bibitem[S\'anchez-Bl\'azquez et al.(2009)]{san09} S\'anchez-Bl\'azquez,
P., Courty, S., Gibson, B. K., \& Brook, C. B. 2009, \mnras, 398, 591

\bibitem[Schaller et al.(1992)]{sch92} Schaller, G., Schaerer, D.,
Meynet, G., \& Maeder, A. 1992, \aaps, 96, 269

\bibitem[Sim\'on-D\'{\i}az(2010)]{sim10} Sim\'on-D\'{\i}az, S. 2010,
\aap, 510, A22

\bibitem[Sim\'on-D\'{\i}az \& Stasinska(2011)]{sim11} Sim\'on-D\'{\i}az,
S.  \& Stasinska, G. 2011, \aap, 526A, 48

\bibitem[Thielemann et al.(1993)]{thi93} Thielemann, F. K., Nomoto,
K., \& Hashimoto, M. 1993, in Origin and Evolution of the Elements,
eds. N. Prantzos, et al., Cambridge University Press, p. 297

\bibitem[Tinsley(1980)]{tin80} Tinsley, B. M. 1980, Fund. Cosmic Phys.,
5, 287

\bibitem[V\'{\i}lchez \& Esteban(1996)]{vil96} V\'{\i}lchez, J. M. \&
Esteban, C. 1996, \mnras, 280, 720

\bibitem[Vlaji\'c et al.(2011)]{vla11} Vlaji\'c, M., Bland-Hawthorn,
J., \& Freeman. K. C. (2011), \apj, accepted (arXiv:1101.0607)

\bibitem[Woosley \& Weaver(1995)]{woo95} Woosley, S. E. \& Weaver, T. A.
1995, ApJS, 101, 181

\bibitem[Yin et al. (2009)]{yin09} Yin, J., Hou, L. K., Prantzos, T. A.,
Boissier, S., Chang, R. X., Shen, S. Y., \&  Zhangh, B. 2009 \aap, 505, 497


\bibitem[Zoccali et al. (2008)]{zoc08} Zoccali, M., Hill, V., Lecureur,
A., Barbuy, B., Renzini, A., Minniti, D., G\'omez, A., \&  Ortolani,
S. 2008, \aap, 86, 177

\bibitem[Zoccali(2010)]{zoc10} Zoccali, M. 2010, in Chemical Abundances
in the Universe: Connecting First Stars to Planets, IAU Symposium 265,
eds. K. Cunha, M. Spite, \& B. Barbuy (Cambridge: Cambridge Univ. Press),
Volume 265, p. 271 \end{thebibliography}
\end{document}